\documentclass[a4paper,11pt]{article}
\pdfoutput=1 

\usepackage{jcappub} 

\usepackage[medium]{titlesec}
\usepackage{subcaption}
\usepackage{graphicx} 
\usepackage{booktabs} 
\usepackage{mathrsfs} 
\usepackage{bm} 
\usepackage{amssymb} 
\usepackage{amsmath}
\usepackage{amsthm}

\usepackage{slashed}
\usepackage{hyperref} 
\usepackage{graphbox} 
\usepackage{physics}
\usepackage{xcolor} 
\usepackage{comment} 
\usepackage{tabularx} 
\usepackage{array} %
\usepackage{multirow} 
\usepackage{ wasysym } 
\hypersetup{
    colorlinks=true,
    linkcolor=blue,
    urlcolor=cyan,
}

\usepackage[hmargin=.7in,vmargin=1.1in]{geometry}
\usepackage[utf8]{inputenc} 
\usepackage{indentfirst}
\graphicspath{{figs/}}

\newcommand{\blue}{\color{blue}}

\newcommand{\black}{\color{black}}

\newcommand{\bge}{\begin{equation}}
\newcommand{\ede}{\end{equation}}
\newcommand{\bga}{\begin{align}}
\newcommand{\eda}{\end{align}}
\def\bgp{\begin{pmatrix}}
\def\edp{\end{pmatrix}}
\def\bgs{\begin{subequations}}
\def\eds{\end{subequations}}

\def\pd{\partial}

\title{Ultra-light dark matter with non-canonical kinetics reopening the mass window}

\author[a,b,d,e]{Shiyun Lu,}
\author[a,b]{Amara Ilyas,}
\author[a,b,c]{Xiao-Han Ma,}
\author[a,b]{Bo Wang,}
\author[a,b,c]{Dongdong Zhang}
\author[a,b]{and Yi-Fu Cai }

\affiliation[a]{Deep Space Exploration Laboratory/School of Physical Sciences,\\
University of Science and Technology of China, Hefei, Anhui 230026, China}
\affiliation[b]{CAS Key Laboratory for Researches in Galaxies and Cosmology/Department of Astronomy,\\
School of Astronomy and Space Science, University of Science and Technology of China, \\
Hefei, Anhui 230026, China}
\affiliation[c]{Kavli Institute for the Physics and Mathematics of the Universe (WPI),\\
UTIAS, The University of Tokyo, Chiba 277-8583, Japan}
\affiliation[d]{Department of Physics, The Hong Kong University of Science and Technology,\\
Clear Water Bay, Kowloon, Hong Kong, P.R.China}
\affiliation[e]{Jockey Club Institute for Advanced Study, The Hong Kong University of Science and Technology,\\
Clear Water Bay, Kowloon, Hong Kong, P.R.China}

\emailAdd{shiyun@ustc.edu.cn}
\emailAdd{aarks@ustc.edu.cn}
\emailAdd{mxh171554@mail.ustc.edu.cn}
\emailAdd{ymwangbo@ustc.edu.cn}
\emailAdd{don@mail.ustc.edu.cn}
\emailAdd{yifucai@ustc.edu.cn}

\abstract{
Fuzzy dark matter (FDM) with mass around $10^{-22}$ eV is viewed as a promising paradigm in understanding the structure formation of the local universe at small scales. 
Recent observations, however, begin to challenge FDM in return. 
We focus on the arguments between the solution to CDM small-scale curiosities and recent observations on matter power spectrum, and find its implication on an earlier formation of small-scale structure. In this article, we propose a scheme of k-ULDM scalar field with a differently-evolving sound speed, thanks to the non-canonical kinetics. With the help of the Dirac-Born-Infeld (DBI) theory, we illustrate to change the behavior of the quantum pressure term countering collapse, therefore change the history of structure growth. We find that it can truly reopen the ULDM mass window closed by the Lyman-$\alpha$ problem. We will discuss such examples in this paper, while more possibilities remain to be explored.  }


\arxivnumber{2407.19521}

\begin{document}

\maketitle
\flushbottom

\newpage
\section{Introduction}\label{sec.intro}
\begin{figure}\centering
	\begin{subfigure}{0.45\textwidth}
	\includegraphics[width=\textwidth]{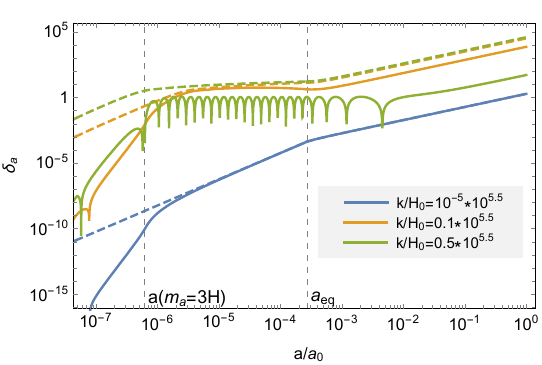}
\caption{
}
\label{fig.FDMda-22}
	\end{subfigure}\qquad
	\begin{subfigure}{0.45\textwidth}
	\includegraphics[width=\textwidth]{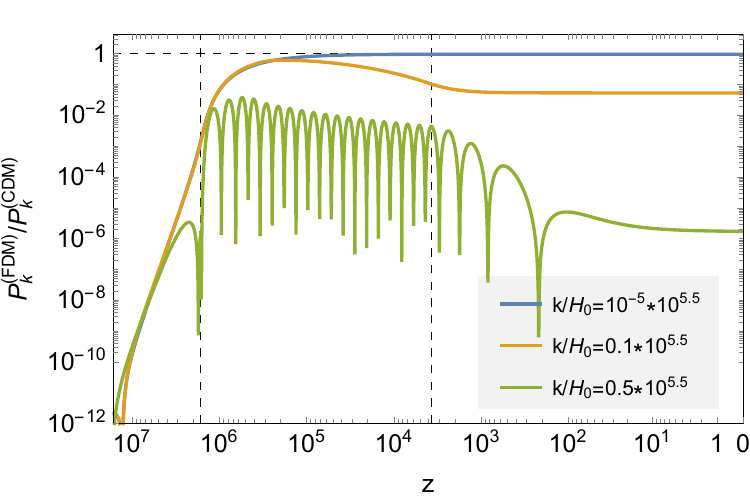}
\caption{
}
\label{fig.Tfdm-22}
	\end{subfigure}
\caption{\textbf{L}eft panel: The comparison of the evolution of DM overdensity, referring to \cite{Marsh:2015xka,Hlozek:2014lca}. The solid lines are for FDM {with} $m_a=10^{-22}\text{ eV}$ constituting all DM, while the dashed lines represent the standard CDM. Note that the structure is suppressed for $k>k_J(a)=66.5 a^{1/4} \text{Mpc}^{-1}$, or {we express it in the unit of Hubble {constant} today as} $k_J(a=a_0)/H_0\simeq 10^{5.5}$. 
\textbf{R}ight panel: The {linear suppression of power spectrum for FDM with $m_a=10^{-22}\text{ eV}$, compared to the fiducial $\Lambda$CDM model, presented by the ratio} $ P^{(\text{\tiny FDM})}_k/P^{(\text{\tiny CDM})}_k$. For the k-modes we concern, the { ratio of the power spectra} is almost a constant at low redshift. 
}
\end{figure}

\begin{figure}
	\centering
	\begin{subfigure}[t]{0.4\linewidth}
	\raisebox{-.52\totalheight}{\includegraphics[width=\textwidth]{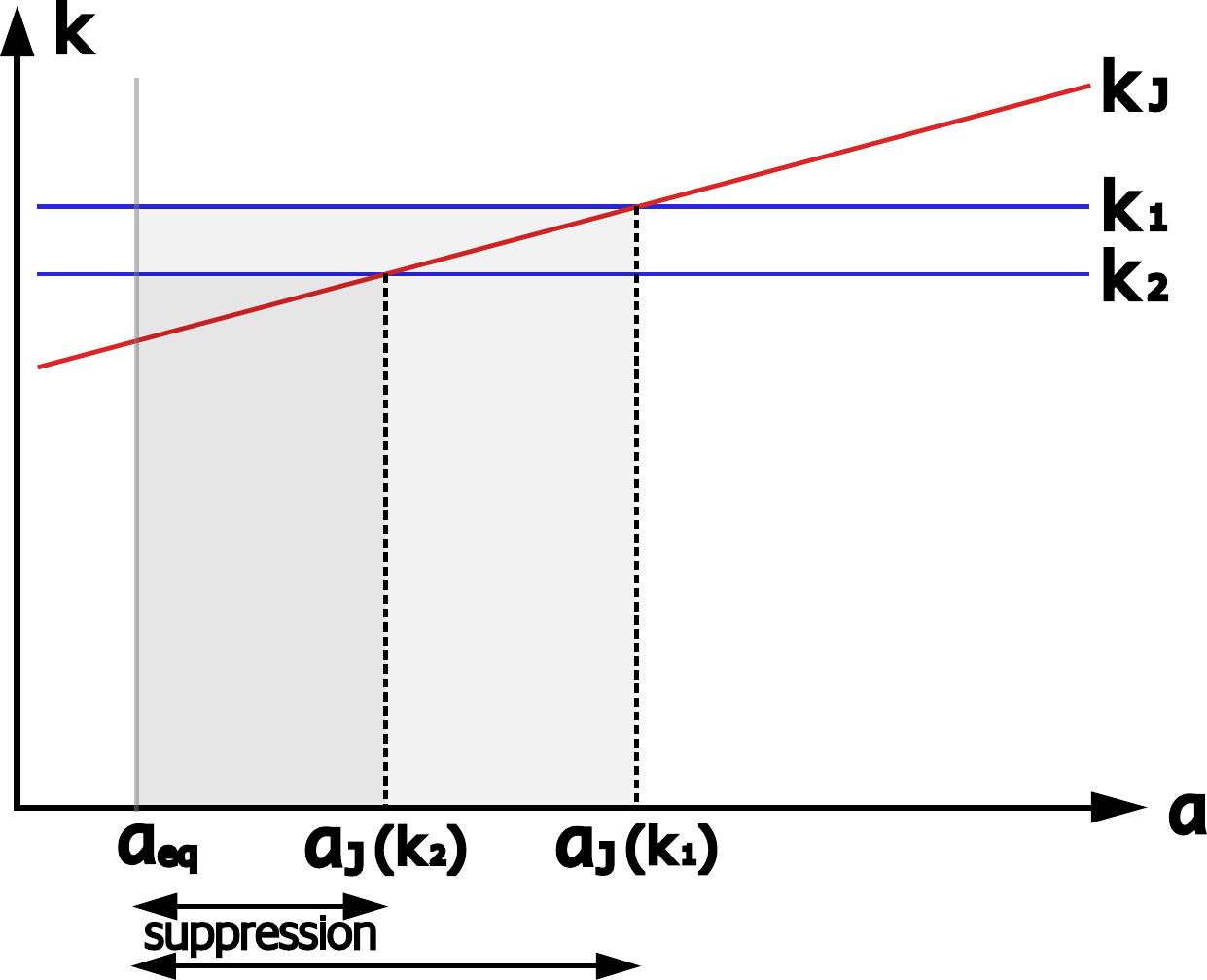}}
\subcaption{}
\label{fig.sup}
	\end{subfigure}\qquad$\Rightarrow$\qquad
	\begin{subfigure}[t]{0.4\linewidth}
	\raisebox{-.49\totalheight}{\includegraphics[width=\textwidth]{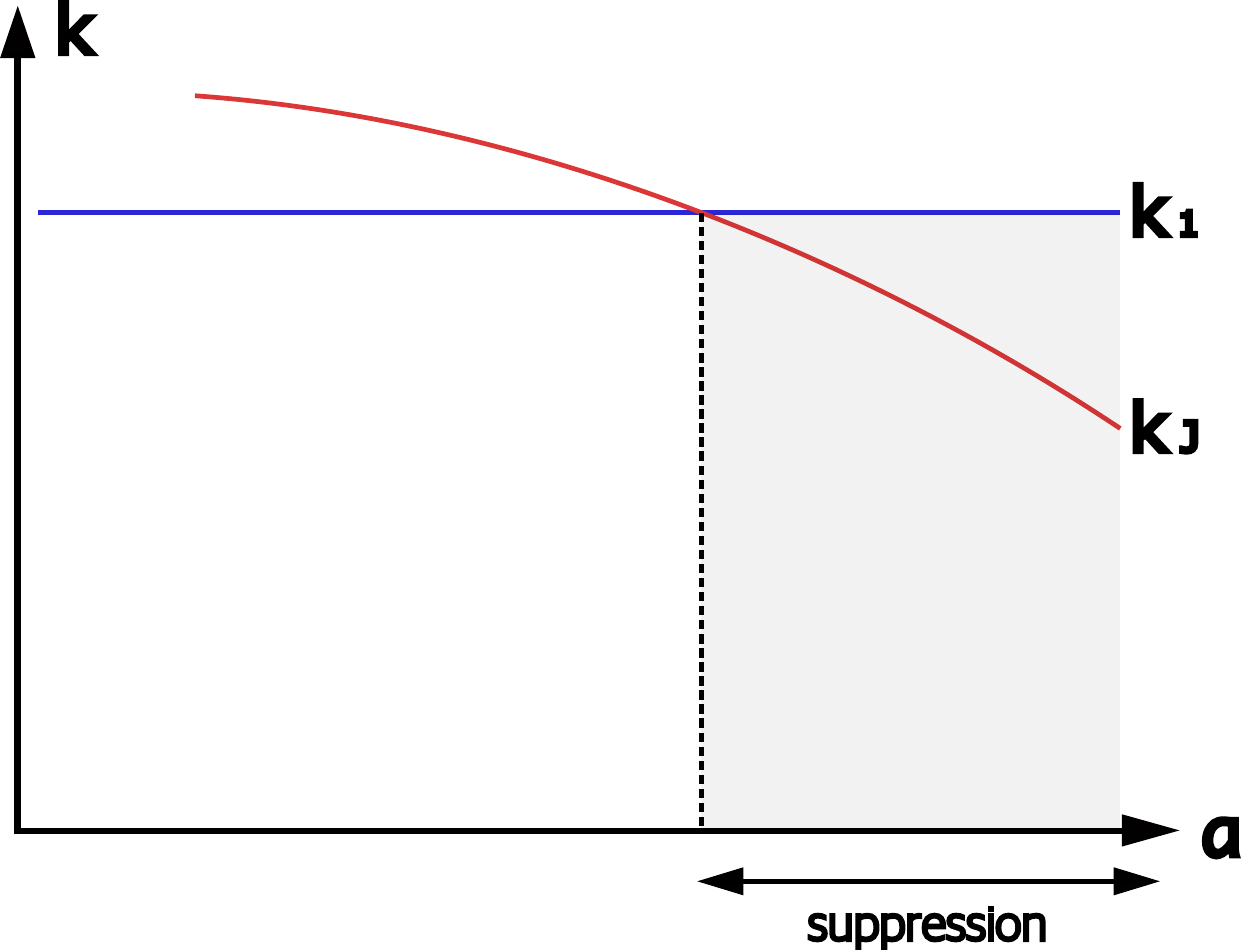}}\\[2.8mm]
\subcaption{}
\label{fig.supdelayed}
	\end{subfigure}
\caption{Comparison of Jeans scales and $k$-modes. \textbf{L}eft panel shows {that} the suppression of FDM {linear structure formation} is integrated {when} $k>k_J$. \textbf{R}ight panel is a {sketch for} how we may {expect} the delayed suppression to {reopen the mass window}, now that observations support CDM at high redshift while a small-scale suppression at low redshift.}
\end{figure}

The standard $\Lambda$CDM model has gained great success, with agreements to most observations. In $\Lambda$CDM, most of the matter in our universe is constituted of cold dark matter (CDM), which is perfect fluid described by equation of state $w=0$ and sound speed $c_s=0$. As the improvement of observations, however, $\Lambda$CDM model meets many challenges \cite{Perivolaropoulos:2021jda}. The small-scale curiosities including cusp-core problem \cite{Oh:2010mc}, missing satellite problem \cite{Bullock:2010uy} (or ``too big to fail'' problem \cite{Boylan-Kolchin:2011qkt,Boylan-Kolchin:2011lmk}) indicate that the nature of dark matter (DM) needs more careful studies.  See a review in \cite{Bullock:2017xww}. DM model that is not so ``cold'' may serve as the solution to the CDM crisis above, by changing the behavior of DM at small scales, while presenting the same properties like CDM at large scales \cite{Bode:2000gq,Hu:2000ke,Blennow:2016gde,Strigari:2006jf}. 
\\
\indent
Ultra-light DM (ULDM) has gained great attention these years, due to its interpretation of CDM { small-scale curiosities} and its well-motivated model. See a review in {\cite{Marsh:2015xka,Ferreira:2020fam,Matos:2023usa,Hui:2021tkt}}. Scalar-type ULDM {\cite{Khlopov:1985fch,Matos:1998vk}} can be constructed from axion models, including axion-like-particles (ALPs) from string theories. (We also call it fuzzy DM (FDM).) Such models can have small enough masses to present wave nature and resist the collapse of DM halos at galaxy scale. Specifically, the soliton solution guarantees that ULDM forms a Bose-Einstein condensate, therefore predicts a core instead of a cusp at the center of halos. Meanwhile, this also suppresses the formation of DM halos that is too small. Therefore, the cusp-core problem and missing satellite problem are solved. A typical mass of $m_a\sim 10^{-22}$ eV corresponding to de Broglie wavelength around $\mathcal{O}(\text{kpc})$ is proper to interpret the observations \cite{Hu:2000ke,Maleki:2020sqn}.
\\ \indent
{R}ecent results on small-scale observations are closing the window for ULDM to serve as the main constitution of DM \cite{Kobayashi:2017jcf,Flitter:2022pzf}, especially for FDM with $m_a\sim 10^{-22}$ eV solving CDM {small-scale curiosities}. Lyman-$\alpha$ (Ly$\alpha$) data from redshift around $z=3\sim 5.4$ prefers CDM to $10^{-22}$ eV FDM, as the latter model predicts too much small-scale suppression \cite{Irsic:2017yje,Kobayashi:2017jcf}. Future observations such as 21-cm {may} further close the parameter window for ULDM \cite{Flitter:2022pzf,Jones:2021mrs}. Besides, subhalo mass function (SHMF) constraints from strong lensing and stellar streams also set lower bound around $m_a\gtrsim 10^{-21}$ eV for FDM~\cite{Schutz:2020jox,Benito:2020avv}. Review of recent observational constraints and analysis can also be found in \cite{Ferreira:2020fam}. 
{As the development of simulation,} more recent researches \cite{Jackson:2023sse} {revisit the cusp-core problem and argue} that, to produce the large cores $r_c\sim\mathcal{O}(100)$ pc {observed} in bright galaxies \cite{Chen:2016unw}, {we do not need such light FDM}, { when} taking the baryonic effects into account. However, { even if} accepting such interpretation, such FDM mass $m_a>10^{-21}$ eV is still in problem with the BHSR-{excluded} range $2.9\times 10^{-21}\text{ eV}<m_a<4.6\times 10^{-21}\text{ eV}$ \cite{Davoudiasl:2019nlo} and $7\times 10^{-20}\text{ eV}<m_a<10^{-16}\text{ eV}$ \cite{Stott:2018opm}. {Besides, the effects of baryonic feedback are still not fully understood \cite{Banares-Hernandez:2023axy,Elgamal:2023yzt}.}
\\ \indent
{With all these considerations, we focus on the FDM mass window around $10^{-21}\sim 10^{-20}$ eV, and notice the redshift subtlety.\footnote{We have to focus on the FDM mass window around $10^{-21}\sim 10^{-20}$ eV, cause a $m_a>10^{-16}$ eV {is not able} to interpret the core size in ultra-faint galaxies \cite{Orkney:2021wmt,Hayashi:2021xxu,Dalal:2022rmp}.} That is, the small-scale observations} prefer {cold enough} DM at high redshift (Ly$\alpha$ at $z=3\sim 5.4$ \cite{Viel:2013fqw,Irsic:2017yje,Kobayashi:2017jcf}) but {require} wave-like DM at low redshift (cusp-core and missing satellite problems from observations of galaxies at $z\approx 0$ {\cite{Ferreira:2020fam}}). {To reopen the mass window constrained by observations at different redshifts, we} construct DM models maintaining the wave nature of ULDM {at $z\approx 0$} but {yielding} different dynamics of structure formation. 
\\ \indent
To achieve this, let us go into details {\blue on} the history of structure formation. The e.o.m of matter overdensity during matter domination epoch (MD) is written as  \cite{Marsh:2015xka,Park:2012ru}
\begin{align}
	\ddot{\delta}_d+2H\dot{\delta}_d+\left(-4\pi G\rho_d+\frac{k^2}{a^2}c_s^2\right)\delta_d=0~,
\end{align}
to linear order, 
where $H=\dot{a}/a$ is the Hubble parameter, and $\delta_d \equiv \delta\rho/\rho$ is the overdensity of DM. The sound speed term $\frac{k^2}{a^2}c_s^2$ competes with gravitational term $4\pi G\rho_d$, and can make $\delta_d$ oscillate instead of growing when it dominates. The sound speed $c_s$ differs FDM from CDM with $c_{s}^{(\text{\tiny CDM})}=0$, as the pressure perturbations in FDM are non-negligible and make an effective $c_{s,\text{eff}}^{2}\equiv\langle\delta p/\delta\rho\rangle_{\text{avg}}=\frac{k^2}{4a^2m^2}/(\frac{k^2}{4a^2m^2}+1)$. For FDM perturbative modes with $k>k_{J}=\mathcal{O}(1)(m_a H_0)^{1/2}a^{1/4}$, the sound speed term dominates and so $|\delta_d|\sim t^0$, resulting in small-scale suppression compared to CDM. {For $k<k_J$, the FDM structure growth becomes indistinguishable from that in CDM. }Figure~\ref{fig.FDMda-22} shows the comparison of structure formation of FDM and CDM, and figure~\ref{fig.sup} shows {the suppression period corresponding to the Jeans scale $k_J(a)$ evolution}. We can see that, as the $k_J$ of FDM grows all the time, the total suppression on k-mode is all accumulated at early time. See figure~\ref{fig.Tfdm-22} for an illustration. Therefore, when comparing the {theoretical power spectrum suppression} $P^{(\text{\tiny FDM})}_k/P^{(\text{\tiny CDM})}_k$ to late time observations, people often neglect its redshift dependence \cite{Hu:2000ke,Irsic:2017yje}.
\\ 
\indent
{To open up the mass window of ULDM, we are expecting a DM model with less wave nature at high redshift compared to that at low redshift, thus} have to reconsider the redshift-dependence of {the power spectrum suppression}. {In other words, we need a modified evolution of $k_J$.} See figure~\ref{fig.supdelayed} for a sketch of $k_J$ to make the suppression not end at very late time. 
From a theoretical aspect, this means a modified sound speed $c_s$. Recall that FDM has introduced non-zero $c_s$ as a solution to CDM {small-scale curiosities}. To extend to DM models with different structure growth behaviors, it is natural to not only make sure the equation of state $w\approx 0$, but reconsider the DM nature by different $c_s$. As DM requires models beyond standard model (SM), we have no reason to reject theories with non-canonical kinetic terms, which {can emerge from string theory} or modified gravities. Such models have theoretically-permitted free parameters to {give non-trivial} $c_s$ {with dynamical evolution behaviors}, and were used for driving inflation or serving as dark energy (DE) \cite{Armendariz-Picon:1999hyi,Garriga:1999vw,Armendariz-Picon:2000ulo,Malquarti:2003nn,Tsujikawa:2010zza,Kobayashi:2019hrl}. But they can also behave like DM (we call it k-DM), as we will focus in this paper. 
Therefore, they can show broad {phenomenons} in changing the history of structure formation, and 
have great potential to {help} understand the small-scale observations, especially in the near future when new data accumulates. 
\\ \indent
{As an example of a framework with non-canonical kinetic terms, the Dirac-Born-Infeld (DBI) theory has been extensively applied in constructing models of inflation \cite{Silverstein:2003hf,Alishahiha:2004eh,Chen:2005ad}, especially with non-trivial sound speeds \cite{Cai:2008if,Cai:2009hw,Cai:2010wt}.}
In this paper, we will discuss the DBI {scalar $\phi$ with kinetic term $f(\phi)^{-1}(1-\sqrt{1+f(\phi)(\pd \phi)^2})$}, and show that DBI DM with $c_s$ controlled by {the wrap factor $f(\phi)$} is truly possible to {meet our expectations.} DBI can model DM not only in the trivial case $f=0$ that goes back to canonical scalar field,  but also at relativistic limit when $c_s^{-1}=(1-f(\phi)\dot{\phi}^2)^{-1/2}\gg 1$.  We will present two typical cases {from} DBI DM. One is FDM-like canonical phase converted from CDM-like non-canonical phase, by assigning $f(\phi)$ with a sudden drop to 0. Another has $c_s$ gradually increasing from a very small {initial} value, making the sound speed term catch up gravitational term at late time.  We will see that {both cases can {give delayed structure suppression and }open up the mass window of ULDM. {DM in the} latter case can {be cold} enough {at high redshift} to meet the Ly$\alpha$ constraints, even when presenting significant wave nature {at $z\approx 0$}. }Such DBI DM model is just one of many possibilities of k-DM, and can be an inspiration of what recent observations may indicate for DM models. 
\\\indent 
Such treatments seem quite rare before, possibly because the discussions of small-scale structure formation history turn to be important only recently, while models from theories with non-canonical kinetics were proposed to drive inflation or serving as DE.
As a remark, our k-ULDM model with increasing $c_s$ is kind of similar to generalized Chaplygin gas \cite{Bento:2002ps}.\footnote{Generalized Chaplygin gas was noticed as a model for unified DM (UDM), serving as DE and DM constitution, by $p= -A\rho^{-\alpha}$ (and so $\rho a^3=(B+Aa^{3(1+\alpha)})^{\frac{1}{1+\alpha}}$). In our case, however, $|p|\propto \rho^{\alpha}$ but $0<p\ll \rho$, and we do not consider the ``DE constitution''. Besides, future fate of {the matter} $\phi$ in our model will also be different from that of Chaplygin gas. } Generalized Chaplygin gas model gained attention as an unified model for constituting most DE and DM (UDM) \cite{Bento:2002ps}, but was excluded then by comparing to observed power spectrum \cite{Sandvik:2002jz}. Further studies on Chaplygin-gas-like UDM noticed their small-scale behavior, and also relate such phenomenological models to theories with non-canonical kinetics \cite{Giannakis:2005kr}. But we do not have to look for a unified model interpreting DE at the same time, especially when trying to understand so many small-scale curiosities of DM. 
Recently, some works have also noticed small-scale structure formation behavior of Chaplygin-gas-like DM. \cite{Suo:2023xes} found a specific model to relieve the $S_8$ tension in the frame of ``generalized DBI''. {There are also researches on self-interacting DM (SIDM) {\cite{Leong:2018opi,Adhikari:2024aff,Leonard:2024mqo}}, which can modify the dispersion relations and potentially solve the small-scale curiosities.} Our work, however, gives a theoretical analysis of the $c_s$ impact on structure formation history, and {indicates} a general theoretical regime with quite rich behaviors. 
We are now expecting more physical interpretation {of} observations at small scales.
\\ \indent
The paper is organized as following. In section~\ref{sec.deltagrow}, we will discuss the treatment of fluid equation of motion from perturbation theories. Then in section~\ref{sec.DBI}, we show DM models that {can} be constructed from DBI theory, {and discuss the sound speed in these models.} 
The analysis of modified sound speed and the results of {k-DM} examples can be found in section~\ref{sec.exp}, with comparison to observational benchmarks. More discussions on perturbation theories and further analysis of sound speed term behavior can be found in appendices. 
\paragraph{Conventions and notations}
In the whole paper we take the convention for metric as $(-,+,+,+)$. The FRW metric is $ds^2=-dt^2+a(t)^2 (dx^i)^2$. The dots overhead or $\left(...\right)^\cdot{}$ represent derivatives with respect to the physical time $t$, and the primes $(...)^{\prime}$ represent derivatives with respect to the conformal time $\tau$. The Hubble parameter is defined by $H\equiv \dot{a}/a$.
\section{Structure growth}\label{sec.deltagrow}
The fluctuations originated from the early universe can seed overdensed and undensed regions. At late time, the overdensities collapse and gradually form the {nearly} homogeneous and isotropic large scale structure, with DM halo and galaxy {structure} at smaller scales. \\ \indent
We can describe such process by the evolution of matter overdensity $\delta\equiv \frac{\delta\rho}{\rho}$. The general e.o.m for fluids can be obtained from the conservation equations $\nabla_\mu T^{\mu\nu}=0$. To the linear order 
\begin{align}\left\{
	\begin{aligned}
	&\dot{\delta}+3H(c_{s,g}^2-w)\delta=-(1+w)(\theta+\dot{h}/2)\\
	&\dot{\theta}+\left[\frac{\dot{w}}{1+w}+(2-3w)H\right]\theta=\frac{k^2}{a^2}\left(\frac{c_{s,g}^2}{1+w}\delta+\Phi\right)
	\end{aligned}~,
\right.
\label{eqt.fluideom}
\end{align}
where $\delta\equiv \frac{\delta\rho}{\rho}$, $\theta\equiv -\frac{k^2}{a^2}\frac{\delta q}{p+\rho}$, $w\equiv p/\rho$, gauge-dependent sound speed defined from $c_{s,g}^2\equiv \frac{\delta p}{\delta \rho}$, and $\dot{h}/2=-3\dot{\Psi}+k^2(B/a-\dot{E})$. Here we have decomposed the energy-momentum tensor into
\begin{align}
	-T_0^0=\rho+\delta\rho~,\quad T_i^i=p+\delta p~,\quad T_i^0=\delta q_{,i}~,
\end{align}
where we have no anisotropic tensor $\Sigma_i^j$ for scalar field, and we have taken the FRW metric perturbations to the first order as 
\begin{align}
	ds^2=-(1+2\Phi)dt^2+2a(t)B_{,i}dx^idt+a(t)^2[(1-2\Psi)\delta_{ij}+2E_{,i|j}]dx^idx^j~.
\end{align}
Note that the precise result of structure evolution should be obtained by numerical simulations. Cause within galaxies, the overdensity can be much larger than $\mathcal{O}(1)$ and the non-linearity should be considered. We use the analytical methods in this paper for relating the theoretical model to structure formation more {explicitly}, and leave the considerations for non-linearity to further work.\footnote{Recent works show it is also possible to treat non-linearity by some analytical methods. See \cite{Li:2018kyk, Kunkel:2022ldl} for calculating by expanding $\delta\rho$ to nonlinear order. See \cite{Taruya:2022zmt} for calculating the FDM halo core structure by spherical mode function methods.} 
\\ \indent
With these, first, let us review how to evaluate the structure growth in fuzzy dark matter (FDM) model. 
After PQ symmetry breaking, the background evolution of a FDM scalar field with mass $m$ is 
\begin{align}
	\ddot{\phi}+3H\dot{\phi}+m^2\phi=0~,
\end{align}
and the equation of state $w\equiv \frac{p}{\rho} = \frac{\dot{\phi}^2/2-m^2\phi^2/2}{\dot{\phi}^2/2+m^2\phi^2/2}$. 
When $m\ll 3H$, $\phi$ rolls down the potential slowly, and behaves like dark energy (DE). While when $H=3m$, $\phi$ starts oscillation and effectively behaves like dark matter (DM), which can be described by an ansatz using WKB approximation~\cite{Marsh:2015xka}
\begin{align}
	\phi(t)=a(t)^{-3/2}(\phi_0\cos(mt+\varphi))~,
\end{align}
where $\phi_0$ and $\varphi$ are constants. 
Typical FDM should have light enough mass $m=10^{-25}\text{ eV}\sim \mathcal{O}(1)$ eV, which can be physically realized by QCD axions or axion-like-particles (ALPs) \cite{Marsh:2015xka,Ferreira:2020fam}. The axion perturbations can be written as
\begin{align}
	\left\{
	\begin{aligned}
	\delta\rho_a&=\dot{\phi}\delta\dot{\phi}-\Phi\dot{\phi}^2+m^2\phi\delta\phi\\
	\delta p_a&=\dot{\phi}\delta\dot{\phi}-\Phi\dot{\phi}^2-m^2\phi\delta\phi\\
	\delta q_a&=-\dot{\phi}\delta\phi
	\end{aligned}
\right.~.
\end{align}
\indent
The axion fluctuations $\delta\phi$ originated from inflation contribute all to isocurvature perturbations, so its adiabatic mode is initially $\delta_a=0$ and $\theta_a=0$ (to zeroth order) \cite{Marsh:2015xka}. The adiabatic mode can grow only when the equation of state of $\phi$ deviates $-1$. 
We can write $\delta p_a=\delta \rho_a+3H(1-\dot{p}/\dot{\rho})\delta q_a$, and rewrite \eqref{eqt.fluideom} as  
\begin{align}
	\left\{
	\begin{aligned}
		\delta_a' &=-k u_a-(1+w_a)h' /2-3aH(1-w_a)\delta_a-9a^2H^2(1-c_{\text{ad}}^2)u_a/k \\
		u_a' &=-aH u_a+k\delta_a+3aH(1+w_a-c_{\text{ad}}^2)u_a
	\end{aligned}
\right.~,
\end{align}
in synchronous gauge, where $u=-\frac{k}{a}\frac{\delta q}{\rho}$ and the adiabatic sound speed $c_{\text{ad}}^2\equiv\frac{\dot{p}}{\dot{\rho}}={w_a}-\frac{\dot{w}_a}{3H(1+w_a)}$. 
The proper initial conditions has been discussed in \cite{Hlozek:2014lca}. 
This set of equations are used for axions before oscillation $a<a_{\text{osc}}$. After $a_{\text{osc}}$, the oscillations make too many steps when we try to solve equations. 
For simplification, we take average of the fast oscillation and get an expression 
\begin{align}
	c_{a}^2=\left\langle \frac{\delta p}{\delta \rho} \right\rangle=
\frac{\frac{k^2}{4m^2a^2}}{1+\frac{k^2}{4m^2a^2}}~,
\end{align}
in axion comoving gauge that $\langle\delta q\rangle=0$ \cite{Park:2012ru}. Transforming to synchronous gauge and use $\langle w_a\rangle=0$, we have 
\begin{align}
	\left\{
	\begin{aligned}
		\delta_a' &=-k u_a-h' /2-3aH c_a^2\delta_a-9a^2H^2c_a^2 u_a/k \\
		u_a' &=-aH u_a+c_a^2k\delta_a+3aHc_a^2u_a
	\end{aligned}
\right.~.
\end{align}
During radiation domination epoch (RD), $\rho_r\gg \rho_m$. To evaluate overdensities $(\delta_a,\delta_r)$ more clearly, we can use 
\begin{align}
	&\ddot{\delta}_r+\frac{1}{2t}\dot{\delta}_r+\left(-\frac{1}{t^2}+\frac{k^2}{3a^2}\right)\delta_r=-\frac{2}{3t}\dot{\delta}_a+\frac{2}{t}\left(c_s^2\delta_a\right)^\cdot{}-\frac{3}{t^2}c_s^2\delta_a~,\\
	&\ddot{\delta}_a+\frac{1}{t}\dot{\delta}_a=\frac{3}{4t^2}\delta_r-\left(\frac{3}{2t^2}+\frac{k^2}{a^2}\right)c_s^2\delta_a~,
\label{eqt.daRD}
\end{align}
from the result of \cite{Park:2012ru} in axion comoving gauge.
During matter domination (MD) when axions constitute most matter $\rho_a\simeq \rho_m \gg \rho_r$, we can also give 
\begin{align}
	\ddot{\delta}_a+2H\dot{\delta}_a+\left(\frac{k^2}{a^2}c_{s,g}^2-4\pi G\rho_a\right)\delta_a=0~,
\end{align}
by combining the equations in \eqref{eqt.fluideom} and use the Poisson equation $\frac{k^2}{a^2}\Psi_B=-4\pi G\delta\rho$.\footnote{This can be obtained from Einstein equations, by taking the subhorizon limit $k\gg aH$. See also in appendix~\ref{sec.apA}.
}
\\ \indent
The behavior of CDM overdensity can also be approximated by FDM with large enough mass, during $a>a_{\text{osc}}(m)$ \cite{Hlozek:2014lca}. For our k-ULDM {(or we sometimes call it k-DM)}, the standard treatment should also start from \eqref{eqt.fluideom}, while the derivations should follow the discussions of perturbations in appendix~\ref{sec.apA}. 
\section{DM model from DBI scalar}\label{sec.DBI}
\indent
DBI theory \cite{Silverstein:2003hf,Alishahiha:2004eh,Chen:2005ad} can change the sound speed of $\phi$ by introducing an arbitrary function of field value $f(\phi)$ into the kinetic term. The action is 
\begin{align}
	S=\int d^4x \sqrt{-g}\left[f(\phi)^{-1}(1-\sqrt{1-2f(\phi)X})-V(\phi)\right]~,
\label{eqt.DBIaction}
\end{align}
where $X=-\frac{1}{2}g^{\mu\nu}\nabla_{\mu}\phi\nabla_{\nu}\phi$. The energy-momentum tensor is 
\begin{align}
	T_{\mu}^{\nu}\equiv-\frac{2g^{\nu\sigma}}{\sqrt{-g}}\frac{\delta(\sqrt{-g}\mathcal{L}_m)}{\delta g^{\mu\sigma}}
=\frac{1}{\sqrt{1-2fX}}\pd_\mu\phi\pd^\nu\phi+\delta_\mu^\nu\left(\frac{1-\sqrt{1-2fX}}{f}-V\right)~.
\end{align}
If we consider only the background, there is $X=\frac{1}{2}\dot{\phi}^2$, then the energy density and pressure can be written as 
\begin{align}\label{eqt.rhodbi}
	\rho&=\frac{1}{c_s}\frac{1}{c_s+1}\dot{\phi}^2+\frac{1}{2}m^2\phi^2~,\\
	p&=\frac{1}{c_s+1}\dot{\phi}^2-\frac{1}{2}m^2\phi^2~,
\end{align}
with the sound speed squared
\begin{align}
	c_s^2\equiv\frac{\pd_{\scriptscriptstyle{X}} p}{\pd_{\scriptscriptstyle{X}} \rho}=1-f(\phi)\dot{\phi}^2~.
\end{align}
The continuity equation $\dot{\rho}+3H(p+\rho)=0$ in {FRW} spacetime gives the equation of motion
\begin{align}
	\ddot{\phi}+3Hc_s^2\dot{\phi}+c_s^3V'(\phi)+\frac{f'(\phi)}{2f(\phi)}\left(1-\frac{2c_s^2}{1+c_s}\right)\dot{\phi}^2=0~.
\label{eqt.dbieom}
\end{align}
\indent
We can infer that when $f\dot{\phi}^2\ll 1$, the sound speed $c_s^2\simeq 1$, and the DBI field $\phi$ turns back to canonical scalar. So, {taking} $V(\phi)=m^2\phi^2/2$, this is the same as the model of fuzzy dark matter (FDM) when the mass is light enough. 
We can find another case when $\phi$ serves as DM at the ``relativistic limit'' $c_s^{-1}=(1-f\dot{\phi}^2)^{-1/2}\gg 1$, that 
\begin{align}
	w=p/\rho\simeq c_s\rightarrow 0~. 
\end{align}
In this case, the sound speed term dominates over mass term in energy density \eqref{eqt.rhodbi}. The relativistic limit can be achieved by fixing $f(\phi)$, which will be discussed in details in section~\ref{sec.exp} and appendix~\ref{sec.apC}. \\
\indent
Although both cases can model the DM, the perturbation evolution can be different because of the $c_s$, and this is exactly what we concern when talking about the structure growth. 
Here let us clarify that the $c_s$ defined here is consistent with the sound speed defined from Mukhanov-Sasaki variables \cite{Sasaki:1986hm,Mukhanov:1988jd}. While when studying the structure growth, people are actually using 
\begin{align}
	c_{s,g}^2\equiv \frac{\delta p}{\delta \rho}~,
\end{align}
which is defined from perturbations and can apparently be gauge-dependent. We have checked in appendix~\ref{sec.apA} that, for DBI scalar when $c_s\rightarrow 0$, 
\begin{align}
	c_{s,g}^2=c_s^2~. 
\end{align}
While for oscillating $\phi$ with $c_s=1$, the effective sound speed 
\begin{align}
	c_{s,g}^2=\left\langle \frac{\delta p}{\delta \rho} \right\rangle=
\frac{\frac{k^2}{4m^2a^2}}{1+\frac{k^2}{4m^2a^2}}~,
\end{align}
in effectively comoving gauge by taking average of fast oscillations \cite{Park:2012ru}. 
{
\\ \indent
The perturbation evolution of DBI scalar fluid can still be described by \eqref{eqt.fluideom}. In the section~\ref{sec.exp}, we will try to construct k-DM examples from DBI, and analytically solve the $\delta_d$ evolution using the methods in section~\ref{sec.deltagrow}, still with the initial conditions following \cite{Hlozek:2014lca}. Our k-DM model is supposed to differ from CDM starting from a late time.  We will see that the main difference of $\delta_d$ evolution for different DM models is from the MD epoch when the structure forms, which is described by \eqref{eqt.dcseomr}. Cause during RD, the evolution of DM overdensity $\delta_d$ almost follows radiation overdensity $\delta_r$, and different $c^2_{s,g}$ has no much effect. 
}
\section{ULDM with modified $c_s$}\label{sec.exp}
Recall the analysis in the introduction. That is, the small-scale curiosities from nearby galaxies and the Ly$\alpha$ data may indicate a DM model behaving like CDM at first and like FDM at late time.
We can construct such DM model by sound speed that is very small at early time and large enough (to encounter collapse) at late time. {Still, we have 
\begin{align}
	\ddot{\delta}_d+2H\dot{\delta}_d+\left(\frac{k^2}{a^2}c_{s,g}^2-4\pi G\rho_d\right)\delta_d=0~,
\label{eqt.dcseomr}
\end{align}
for the overdensity $\delta_d\equiv \delta\rho_d/\rho_d$ during MD epoch, and the {growth} of $\delta_d$ is determined by the dominance of the quantum pressure term $\frac{k^2}{a^2}c_{s,g}^2$ or the gravitational term $4\pi G\rho_d$, as discussed in the introduction. 
To see how the modified $c_s$ change the $\delta_d$ behavior more specifically, we also discuss some toy models 
in appendix~\ref{sec.apB}.}
\\ \indent
DM model with ultralight mass and different sound speed can be realized by scalar field with non-canonical kinetic terms, including the k-essence and other EFTs. We follow such idea and construct examples from DBI theory, and show that it is truly possible to make new DM model compatible with Ly-$\alpha$ observations, while still have enough small-scale suppression to solve cusp-core problems. Note that to keep the enough total suppression {at $z\approx 0$}, the mass of the new DM can be different from that of FDM.
\subsection{Example 1: ``phase transition'' case (late time suppression by switching to FDM-like phase)}\label{sec.eg1}
A natural idea to interpret Ly$\alpha$ by new ULDM model is to make it {change from} CDM-like {to} FDM-like {at} some particular time $t_c$. We call such case ``phase transition'', since it requires a sudden jump of $f(\phi)$ between two totally different phases for DBI DM, as we will see later. 
\subsubsection{Background evolution}
\begin{figure}[t]\centering
	\begin{subfigure}{0.45\textwidth}
	\includegraphics[width=\textwidth]{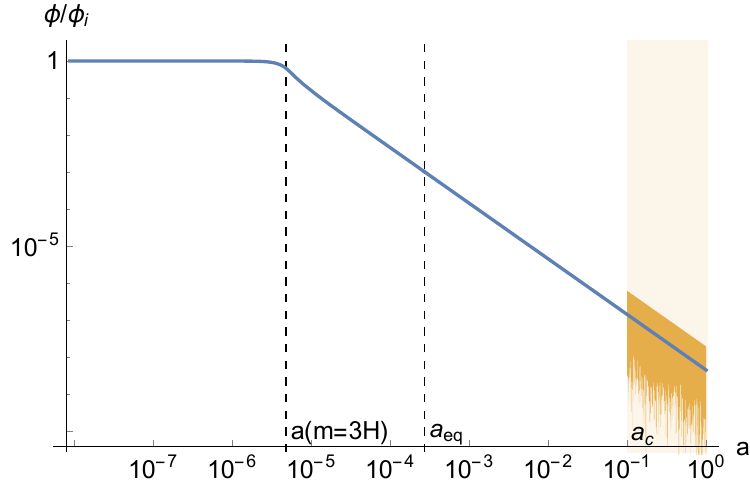}
\caption{The evolution of field value of $\phi$. 
} \label{fig.phic}
	\end{subfigure}\quad
	\begin{subfigure}{0.45\textwidth}
	\includegraphics[width=\textwidth]{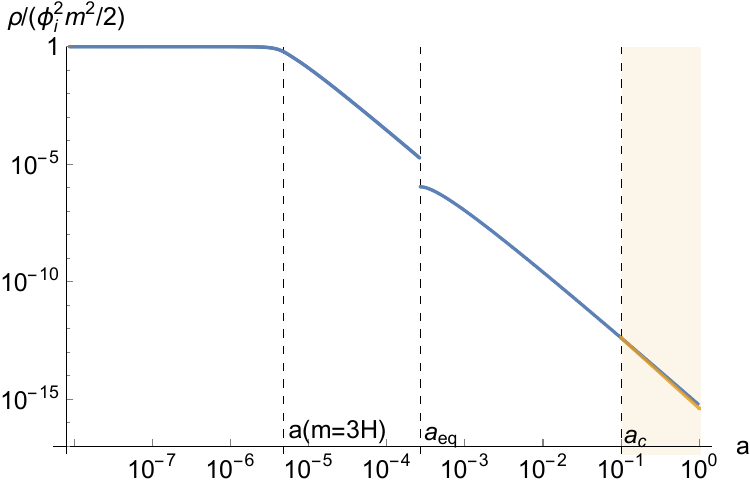}
\caption{The evolution of the energy density $\rho$ of $\phi$. 
} \label{fig.rhoc}
	\end{subfigure}
\caption{The evolution of $\phi$, from $t_i=10^{-5}t_{\text{osc}}$ till $t_0$, by taking the $f(\phi)$ in \eqref{eqt.fphi} with $m=1.6\times 10^{-24}\text{ eV}$ and $a_c=0.1$. The blue lines represent the CDM-like behavior of $\phi$, while the yellow lines for the FDM-like $\phi$, with the epoch of switched-off $f(\phi)$ shown by the light yellow regions. The vertical lines are the {scale factors at} $t_{\text{osc}}$ when oscillation should have begun for FDM when $H\simeq m/3$, at equality $t_{\text{eq}}$, {and at switching time $t_c$}, respectively. 
\textbf{L}eft panel: The evolution of the background value of $\phi$. \textbf{R}ight panel: The evolution of the energy density $\rho$ of $\phi$. {The sudden drop at $t_{\text{eq}}$ probably comes from the inconsistency of $H(t)$ at $t_{\text{eq}}$, {as} we naively treat the transition from RD ($H=\frac{1}{2t}$) to MD ($H=\frac{2}{3t}$) as a sudden transition. 
\black}
}
\label{fig.fDM}
\end{figure}

\begin{figure}\centering
	\includegraphics[width=0.48\textwidth]{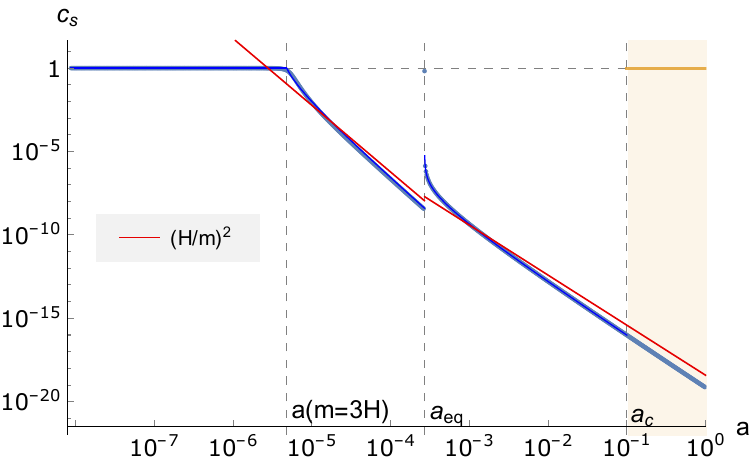}
	\includegraphics[width=0.47\textwidth]{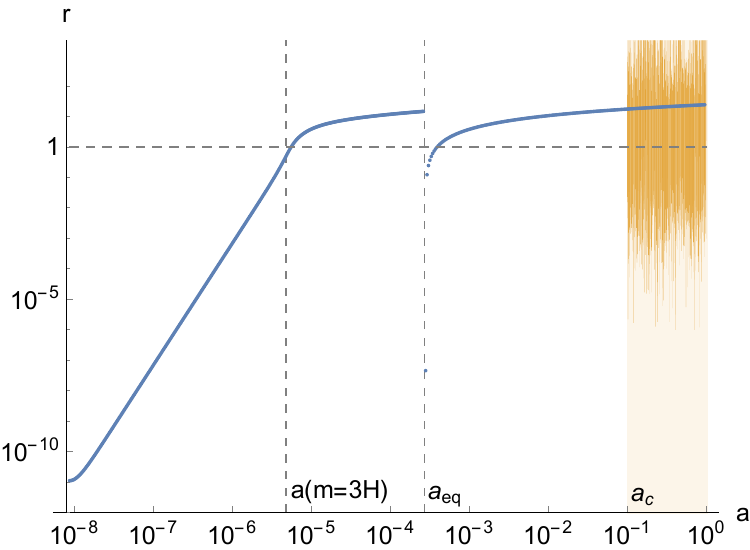}
\caption{The evolution of sound speed $c_s$ and the ratio $r$, by taking the $f(\phi)$ in \eqref{eqt.fphi} with $m=1.6\times 10^{-24}\text{ eV}$ and $a_c=0.1$. Same as before, the blue lines represent the CDM-like behavior of $\phi$, while the yellow lines for the FDM-like $\phi$, with the epoch of switched-off $f(\phi)$ shown by the light yellow regions. \textbf{L}eft panel: The evolution of $c_s$ {comparing} to $(H/m)^2$. The bright blue line is the analytical $c_s$ from \eqref{eqt.csiapp}, comparing to numerical lines {in light blue}. \textbf{R}ight panel: The ratio of kinetic term to mass term in $\rho_\phi$, showing the reliability of $c_s$ term dominance. Comparison of $r$ and $1$ is equivalent to the comparison of $c_s$ and $(H/m)^2$. }
\label{fig.csterm}
\end{figure}

To make our DM model to behave like CDM before some $t_c$ and like FDM after $t_c$, we try to set $c_s^2=0$ for $t_{\text{osc}}<t<t_c$, while set $c_s^2$ back to $1$ for $t_c<t<t_0$. We can achieve this by\footnote{The $f(\phi)$ we used here is a function only dependent on $\phi$, so the division points (e.g.~$t_{\text{eq}}$) of this piecewise function should actually be determined by $\phi$. This is still reasonable before $t_c$, when $\phi$ is monotonically decreasing with time as in figure~\ref{fig.phic}, but the determination of $t_c$ is left unclear. The issue may be {improved} by $c_s$ with explicit dependence on $t$ (e.g. through Ricci scalar $R$), {from} scalar-tensor theory \cite{Tsujikawa:2010zza,Kobayashi:2019hrl}.}
\begin{align}
	f(\phi)=\left\{
	\begin{aligned}
		&\frac{1}{(m/H_0)^2}t_0^{2}\phi^{-2}(\phi/\phi_i)^{-8/3}\  &\text{for}\ t_i<t<t_{\text{eq}}~\\ 
		&\left(\frac{4 m}{3H_0}\right)^{-3/2} (t_{\text{eq}}/t_0)^{1/2}t_0^{2} \phi^{-2} (\phi/\phi_i)^{-2} &\text{for}\ t_{\text{eq}}<t<t_c~\\
		&0 
\black &\text{for}\  t>t_c\hspace{28pt}~
	\end{aligned}
\right.~,
\label{eqt.fphi}
\end{align}
where $\phi_i=\phi(t_i)$ when the field is slow-rolling and behaves like DE, and $\phi_c=\phi(t_c)$ when the CDM-like field turns to FDM-like. By solving the e.o.m \eqref{eqt.dbieom} numerically, we can see that, 
\begin{itemize}
	\item Before $t_c$, such $f(\phi)$ 
makes $\rho a^3$ nearly constant without oscillation, and $c_s\ll 1$. The oscillation is switched off by $f(\phi)$, as the kinetic term dominate over the mass term. We show this in figure~\ref{fig.fDM}, by taking $m\sim 
10^{-24} \text{ eV}$ and $a(t_c)=0.1$. 
	\item After $t_c$, as long as the small enough $f(\phi)$ can keep $f(\phi)\dot{\phi}^2\ll 1$ when the oscillation resumes, the CDM-like field $\phi$ turn to be like FDM afterwards. 
\end{itemize}
\paragraph{Some analysis:} For CDM-like $\phi$, we should have $0<p\ll \rho$ (so that $\rho\propto a^{-3}$ can be satisfied, according to the continuity equation) and $c_s\ll 1$, which means that $m^2\phi^2\ll \frac{1}{c_s}\dot{\phi}^2$. We can also check this in figure~\ref{fig.csterm}. 
\ \\ \\
\small
We can analyze the time dependence of $\rho a^3$ as following. The e.o.m \eqref{eqt.dbieom} is equivalent to the continuity equation of background $\dot{\rho}+3H(p+\rho)=0$ as 
$
	\frac{d}{dt}(\frac{1}{c_s+1}\frac{1}{c_s}\dot{\phi}^2+\frac{1}{2}m^2\phi^2)+3H\frac{1}{c_s}\dot{\phi}^2=0~,
$
which can be written as 
\begin{align}
	a^{-3}\left(\frac{1}{c_s}\dot{\phi}^2a^3\right)^\cdot{}+\left(-\frac{1}{c_s+1}\dot{\phi}^2+\frac{1}{2}m^2\phi^2\right)^\cdot{}=0~. 
\label{eqt.conserve}
\end{align}
When $t_{\text{osc}}<t<t_c$, we have $|\dot{\phi}|=\frac{3}{2}H\phi\ll m\phi$, so $\dot{\phi}^2\ll m^2\phi^2$. Defining a ratio of energy density terms $r\equiv \frac{\dot{\phi}^2/(c_s(c_s+1))}{m^2\phi^2/2}$, we can see that $c_s<1$ and $r^{-1}<1$ after $t_{\text{osc}}$, and they quickly decrease, so 
$
	a^{-3}\left(\frac{1}{c_s}\dot{\phi}^2a^3\right)^\cdot{}\simeq -\left(\frac{1}{2}m^2\phi^2\right)^\cdot{}=3H\cdot\frac{1}{2}m^2\phi^2\propto Ha^{-3}
$. 
It means 
\begin{align}
	\rho\ a^3\simeq \frac{1}{c_s}\dot{\phi}^2a^3\propto \log({ A}~t)~,\quad \text{when}\ t_{\text{osc}}<t<t_c~,
\end{align}
where $A$ is a constant, 
and we can see $r\propto \log({ A}\ t)$, as shown in figure~\ref{fig.csterm}. From \eqref{eqt.conserve}, we can also give the analytical expressions for sound speed 
\begin{align}
	c_s=\left\{
	\begin{aligned}
		& \qquad \qquad \qquad \qquad \quad 1 &\text{ for } t<t_{\text{osc}}\text{ or } t>t_c\\
		&\left.\left(\frac{3}{2}H(t)\right)^{2}\left[\left(\frac{3}{2}H(t_{\text{osc}})\right)^2+\frac{3}{4}m^2\log(t/t_{\text{osc}})\right]^{-1} \right.  &\text{ for } t_{\text{osc}}<t<t_{\text{eq}}\hspace{15pt}\\
		&\left.\left(\frac{3}{2}H(t)\right)^{2}\left[c_{s,\text{eq}}^{-1}\left(\frac{3}{2}H(t_\text{eq})\right)^2+m^2\log(t/t_{\text{eq}})\right]^{-1} \right.  &\text{ for } t_{\text{eq}}<t<t_{c} \hspace{22pt}
	\end{aligned}
\right.~,
\label{eqt.csiapp}
\end{align}
by ignoring the $-\frac{1}{c_s+1}\dot{\phi}^2$ term under the condition of $H\ll m$ and $c_s<1$. We can see the analytical form fit well with numerical results in the left panel of figure~\ref{fig.csterm}. The logarithm is from the unignorable mass term, which can be inferred from the comparison of $r$ and $1$, or $(H/m)^2$ and $c_s$. 
For more discussions of $c_s$ and $(H/ m)^2$, see appendix~\ref{sec.apC}.
\normalsize
\subsubsection{Comparing to observation}
\begin{figure}\centering
\begin{subfigure}{0.45\textwidth}
	\includegraphics[width=\textwidth]{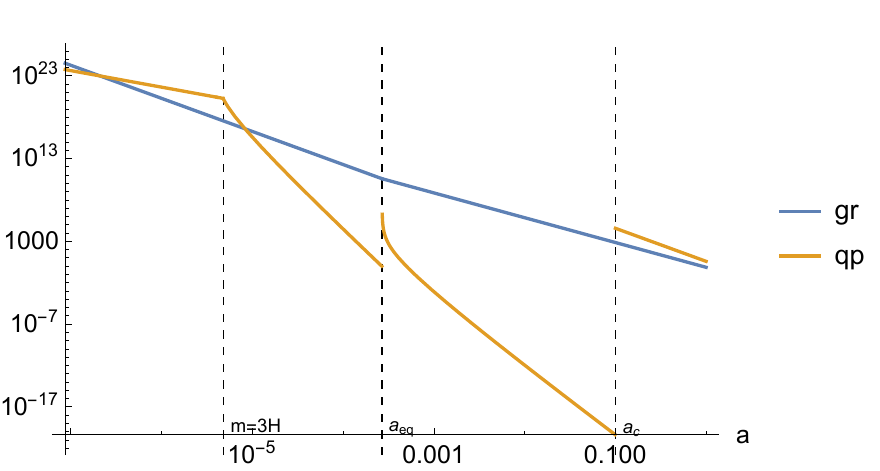}
\caption{The comparison of terms for $k/H_0=3\times10^{4.5}$. 
}
\end{subfigure}\qquad
\begin{subfigure}{0.45\textwidth}
	\includegraphics[width=\textwidth]{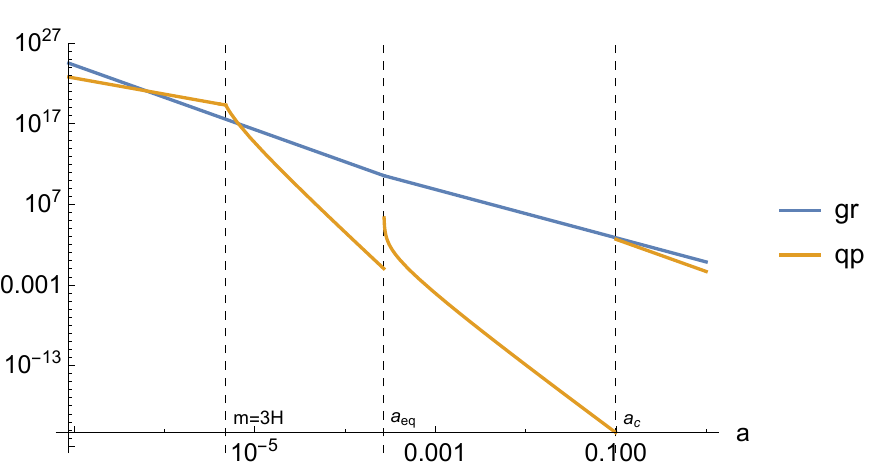}
\caption{The comparison of terms for $k/H_0=1\times10^{4.5}$.}
\end{subfigure}
\caption{A rough comparison of the structure-formation-determining terms in \eqref{eqt.dcseomr}, shown in blues lines for gravitational terms $4\pi G\rho_d$ and orange lines for sound speed terms $\frac{k^2}{a^2}c_{s,g}^2$ (or we call it ``quantum pressure'' as in FDM). The sound speed term counters the collapse and suppresses the structure formation when it becomes dominant during MD. We take the $f(\phi)$ proposed in \eqref{eqt.fphi}, with $a_c=0.1$ and $m=1.6\times 10^{-24}\text{ eV}$ ($m/H_0=1.6\times 10^{9}$). 
}
\label{fig.k3csj}
\end{figure}

\begin{figure}\centering
	\includegraphics[width=0.5\textwidth]{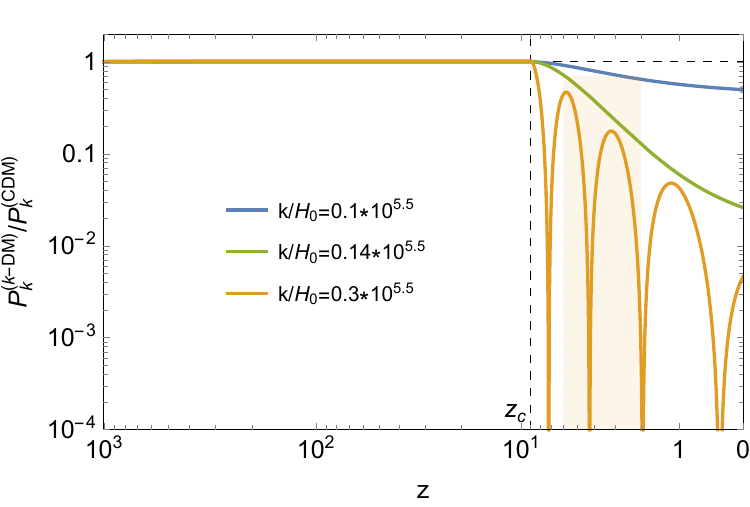}
\caption{{After $t_{\text{eq}}$, }the suppression of overdensity spectrum (compared to that of CDM) 
$P^{(\text{\tiny k-DM})}_k/P^{(\text{\tiny CDM})}_k$ as a function of redshift $z$, for the $f(\phi)$ proposed in \eqref{eqt.fphi} with $a_c=10^{-1}$ and $m_\phi=1.6\times 10^{-24}$ eV. The isolated blue point at $z=0$ is for {the expected wave nature presented by the k-ULDM, with a half mode} $k=0.1\times 10^{5.5}H_0\sim 4.5\text{ Mpc}^{-1}$ {as in \eqref{eqt.half} (corresponding to $10^{-22}$ \text{eV} FDM).} The shaded region is the benchmark from the Ly-$\alpha$, which should lay below the yellow line for $k=0.3\times 10^{5.5}H_0\sim 20h\text{ Mpc}^{-1}${, as required by \eqref{eqt.lya}}. 
We may compare this to figure~\ref{fig.Tfdm-22}, and find that the {inconsistency between} Ly-$\alpha$ {results and late-time wave nature} can be alleviated, compared to that in $m_a=10^{-22}$ eV FDM model.  
}
\label{fig.supcsj}
\end{figure}

Now we check the sturcture evolution of the new k-ULDM model with modified sound speed from DBI theory. 
\\ \indent
For the case with $f(\phi)$ proposed in \eqref{eqt.fphi}, the sound speed decreases almost proportional to $t^{-2}$ after $t>t_{\text{osc}}(m=3H)$, as shown in figure~\ref{fig.csterm}. $c_{s,g}^2=c_s^2$ when $c_s\ll 1$. After $t_c$, the sound speed becomes FDM-like. So the gauge-dependent sound speed is 
\begin{align}
 	c_{s,g}^2=\left\{
	\begin{aligned}
		&\hspace{20pt}1	&t<t_{\text osc}\\
		&\hspace{0pt}\sim(t/t_{\text osc})^{-4}\ &t_{\text osc}<t<t_c\hspace{8pt}\\
		&\hspace{2pt}\frac{\frac{k^2}{4a^2m^2}}{1+\frac{k^2}{4a^2m^2}}	&t>t_c\hspace{8pt}
	\end{aligned}
\right.~,
\end{align}
determining the structure growth. 
So we can roughly compare the gravitational term and sound speed term in \eqref{eqt.dcseomr}, shown in figure~\ref{fig.k3csj}.\footnote{The comparison of sound speed term and gravitational term is not exact, just on the level of ``same order''. }$^{,}$\footnote{During RD, the gravitational term is actually dominated by $8\pi G\rho_r\delta_r${\cite{Park:2012ru}}, and we estimate it by $\delta_r\approx \delta_a$ here.} 
{The structure formation at scale $k$ is suppressed when sound speed term becomes dominant. See} the {power spectrum compared to that of CDM} {$P^{(\text{\tiny k-DM})}_k/P^{(\text{\tiny CDM})}_k$ in figure~\ref{fig.supcsj}}.\footnote{
People usually separate the growth factor $D_+(z)$ to describe the structure growth, and collect other evolution properties into the ``transfer function'' $T(k)$ \cite{dodelson2020modern}. So a late time FDM power spectrum $P_F(k,z)=T_F^2(k)P_{\Lambda{\text{CDM}}}(k,z)=T_F^2(k)[D_+(z)/D_+(0)]^2 P_{\Lambda{\text{CDM}}}(k)$ \cite{Marsh:2015xka}. In our work, we do not fix to these concepts, but directly solve our k-DM overdensity $\delta_d$ evolution starting from $t_i$, and compare the linear power spectrum to fiducial $\Lambda$CDM, with $P^{(\text{\tiny k-DM})}_k/P^{(\text{\tiny CDM})}_k=(\delta_{d}^{(\text{\tiny k-DM})}/\delta_{d}^{(\text{\tiny CDM})})^2$. 
} 
{Taking} the turning point $t_c$ at $a(t_c)=10^{-1}$, we can delay the suppression to a late time. 
{A typical mass for FDM to form a core in DM halo is around $10^{-22}$ eV, with a half mode $k^{(\text {FDM},m_a=10^{-22}\text{ eV})}_{1/2}= 4.5\text{ Mpc}^{-1}\sim 10^{4.5}H_0$ \cite{Kamionkowski:1999vp, Hu:2000ke,Marsh:2015xka}. As an illustration, let our k-DM mimic such FDM at low redshift, with the same half-mode $k_{1/2}^{\text{(k-DM)}}=4.5\text{ Mpc}^{-1}$.} Our k-DM is CDM-like before $t_c$, so we need the $m_{\phi}<10^{-22}$ eV to reproduce the total suppression, where we have taken $m_{\phi}=1.6\times 10^{-24}$ eV in {the figures}. 
{To show the {capability} of our new k-DM model, we pick out the following observations {that} used to constrain the FDM mass window.}
Now we can {check our results with such observations}, which can be simplified to some benchmark\footnote{Ly$\alpha$ constriants from XQ-100 and HIRES/MIKES data, with the measurement at redshift $3\sim 5.4$. Here we use a simplified benchmark referring to \cite{Carena:2021bqm}.}
\begin{align}
	&{P^{(\text{obs})}_k/P^{(\text{\tiny CDM})}_k}{\big |}_{k=4.5\ \text{Mpc}^{-1}}\sim 0.5 \quad \text{cusp-core \& missing satellite, at } z\approx 0 \text{\ (wave nature)} \label{eqt.half}~,
\\ \label{eqt.lya}
	&{P^{(\text{obs})}_k/P^{(\text{\tiny CDM})}_k}{\big |}_{k=20h\text{ Mpc}^{-1}}>0.7\quad \text{Lyman-}\alpha \text{, at } z\approx{ 2\sim 6} \text{\ (CDM-like)}~.
\end{align}
We can see in figure~\ref{fig.supcsj} that, the k-ULDM model from DBI with $f(\phi)$ in \eqref{eqt.fphi} is able to alleviate the mass constraints by Ly$\alpha$, compared to that in FDM (see figure~\ref{fig.Tfdm-22}). 
{{\paragraph{Remarks} Recent simulations \cite{Hayashi:2021xxu,Jackson:2023sse} argue that observations of galaxy cores and Ly$\alpha$ may actually agree on the higher lower bound of FDM mass $m_a>\mathcal{O}(10^{-21})$ eV, getting close to the mass range not allowed by black hole superradiance (BHSR). Nonetheless, the discussion in this section is an illustration to show how our k-ULDM may open up the mass window. At higher redshift, the k-DM is more like CDM, so has better compatibility to higher redshift observations (e.g. Ly$\alpha$) than its corresponding FDM, which can be figured out in figure~\ref{fig.supcsj} and figure~\ref{fig.Tfdm-22}. Therefore, it can alleviate the ``galaxy core - Ly$\alpha$ flux power spectrum problem'' (if it is still a problem for canonical FDM).
Furthermore, to mimic the wave nature of FDM at $z\approx 0$, the k-DM should have $m_\phi< m_a$ to reproduce the total suppression. Whatever the FDM mass lower bound from recent small-scale observational studies, the k-DM here can open up its mass window largely, pushing it away from BHSR-{excluded} mass range. }\\ \indent
Figure~\ref{fig.supcsj} is an illustration of our k-DM to show how it can present significant wave nature at $z\approx 0$ while be close to CDM at higher redshift. 
Note that recent researches on galaxy cores almost agree on DM with less wave nature, corresponding to heavier FDM, or a heavier k-DM. Under such revision, a realistic k-DM can be more like CDM, thus actually more consistent with Ly$\alpha$ than shown in figure~\ref{fig.supcsj}.}
\subsection{Example 2: ``chaplygin-like'' case (late time suppression by increasing $c_s$)}\label{sec.eg2}
The above case in section~\ref{sec.eg1} requires a hand-set sudden switch at $t_c$, and the late time suppression is not quick enough to be compatible to Ly$\alpha$ benchmark. So we show another example as following. A growing $c_s^2$ can make the $k^2c_s^2/a^2$ naturally exceed the gravitational term at late time, with a quicker suppression of $\delta_k$ and different $k$-dependence. Compared to the model above (which has a $c_s$ with ``sudden jump''), such model may better fit the observation. 
\subsubsection{Background evolution}
\begin{figure}\centering
	\includegraphics[width=0.45\textwidth]{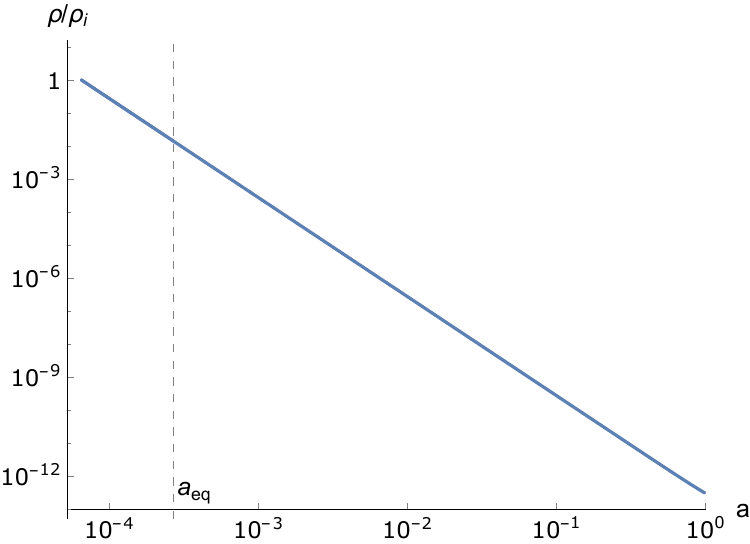}\qquad
	\includegraphics[width=0.45\textwidth]{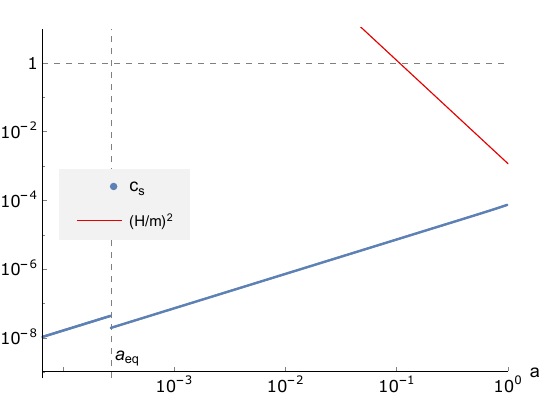}
\caption{The evolution of energy density and sound speed of k-ULDM in {Example 2}, by taking the $f(\phi)$ in \eqref{eqt.fphii}, with $c_{s,i}=c_s(a_i=6.5\times 10^{-5})=1.09\times 10^{-8}$ and $m/H_0=29.0$. {\bf{L}}eft panel: The evolution of $\rho\propto a^{-3}$ just as DM. 
{\bf{R}}ight panel: The evolution of $c_s\propto a$ and its comparison to $(H/m)^2$. Again, the inconsistency {at $t_{\text{eq}}$} comes from the treatment with RD-MD sudden transition. }
\label{fig.csis}
\end{figure}

As an example, let us take 
\begin{align}
	f(\phi)=\left\{\begin{aligned}
	&\left(1-c_{s,i}^2(\phi/\phi_i)^2\right)\left(2t_i\phi_i^{-1}(\phi/\phi_i)\right)^2 
&\text{for } t_i<t<t_{\text{eq}} \text{ during RD}\\
	&\left(1-c_{s,\text{eq}}^2(\phi/\phi_{\text{eq}})^4\right)\left(3t_{\text{eq}}\phi_{\text{eq}}^{-1}(\phi/\phi_{\text{eq}})^2\right)^2 
 &\text{for } t_{\text{eq}}<t<t_c \text{ during MD}
\end{aligned}
\right.~,
\label{eqt.fphii}
\end{align}
where $c_{s,i}$ is very small at the initial time, and $t_c$ is some time (possible in future) when $c_s$ gets close to $(H/m)^2$ and the expression fails. 
So the sound speed 
\begin{align}
	c_s=c_{s,i}(a/a_i)\quad \text{for } t>t_i~,
\end{align}
as long as $c_s\ll (H/m)^2$.\footnote{For the evolution when $c_s$ gets close to $(H/m)^2$, see appendix~\ref{sec.apC}.} Then the sound speed term $\frac{1}{c_s}\dot{\phi}^2\gg m^2\phi^2$, and $\rho\sim\frac{1}{c_s}\dot{\phi}^2\propto a^{-3}$. Such k-ULDM may also serve as DM, with a growing $c_s^2$. 
Figure~\ref{fig.csis} shows the background evolution. Actually this model is kind of like generalized Chaplygin gas model, in the sense of $p=A \rho^{2/3}$ (during MD), and so gives a time-dependent $c_s$ with power law. But our model differs from past Chaplygin gas models that we are only interested in $\phi$ serving as DM, and the coefficient $0<A\ll 1$. 
\\ \indent
So when the $c_s$ gets large enough (but still much smaller than 1), sound speed term dominates the evolution of $\delta$, and starts the suppression {accumulating} more quickly than in FDM case.\footnote{We can surely expect the sound speed term dominates when $c_s$ reaches $\mathcal{O}(1)$. In such condition, however, the DBI $\phi$ goes back to canonical scalar. Then the model is almost same as FDM, and the $\delta$ evolution is similar to the case discussed in section~\ref{sec.eg1}. More discussions about the evolution of sound speed close to $\mathcal{O}(1)$ can be found in appendix~\ref{sec.apC}.} 
We can see this in figure~\ref{fig.k3csi}.  
\paragraph{Remarks}Still, the mass of $\phi$ should be light enough to guarantee the large enough $\frac{k^2}{a^2}c_{s,g}^2$. We can see this by the following analysis. If $m$ is heavy, the large enough $c_s$ have to  exceed $(H/m)^2$ before the late time. According to the appendix \ref{sec.apC}, the $c_s$ quickly jump to $1$ and so mass term dominates $\rho$, our $\phi$ becomes FDM-like. While even for FDM, the mass have to be small enough to present enough wave nature. 
Note that for a small enough $m_\phi$ with $c_s$ getting back to $1$ before $t_0$, $\phi$ can turns to be DE. Such case is closer to unified dark matter (UDM) and also interesting to be checked, but is out of the main concern of this paper. 
\subsubsection{Comparing to observation}
\begin{figure}\centering
\begin{subfigure}{0.45\textwidth}
	\includegraphics[width=\textwidth]{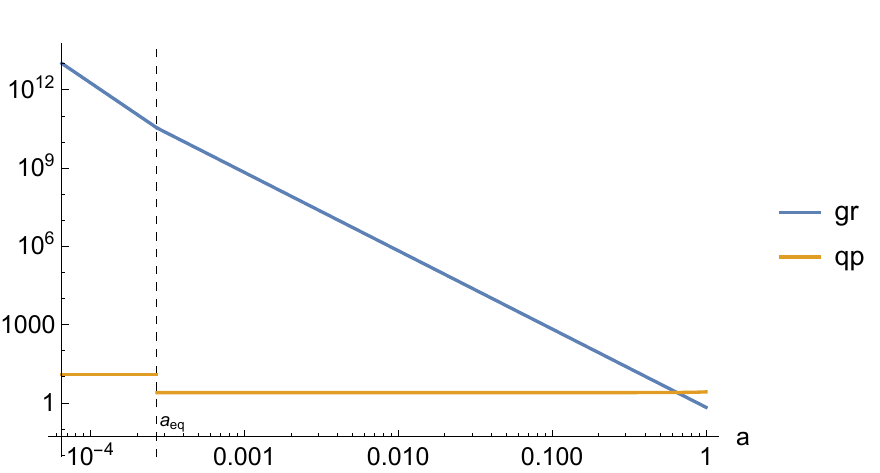}\\
\caption{The comparison of terms for $k/H_0=1\times10^{4.5}$.\\\ \linespread{0pt} 
\\ }
\end{subfigure}
\quad
\begin{subfigure}{0.4\textwidth}
	\includegraphics[width=\textwidth]{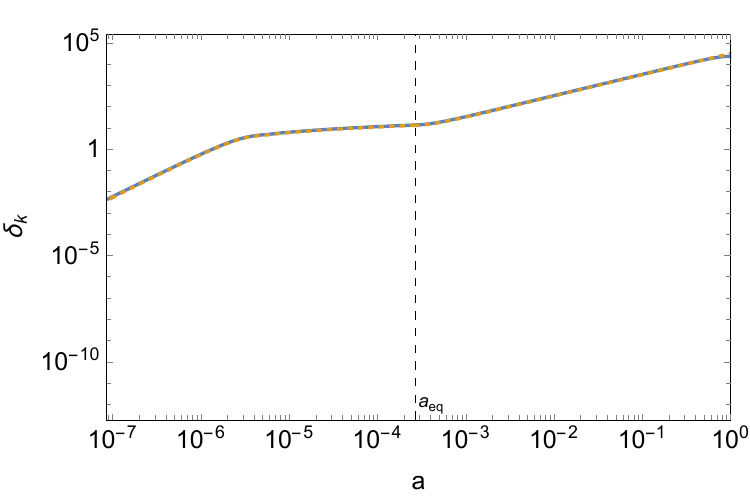}
\caption{$\delta_k$ evolution for our k-ULDM (solid line) and CDM (dashed line), with $k/H_0=1\times10^{4.5}$.}
\end{subfigure}
\begin{subfigure}{0.455\textwidth}
	\includegraphics[width=\textwidth]{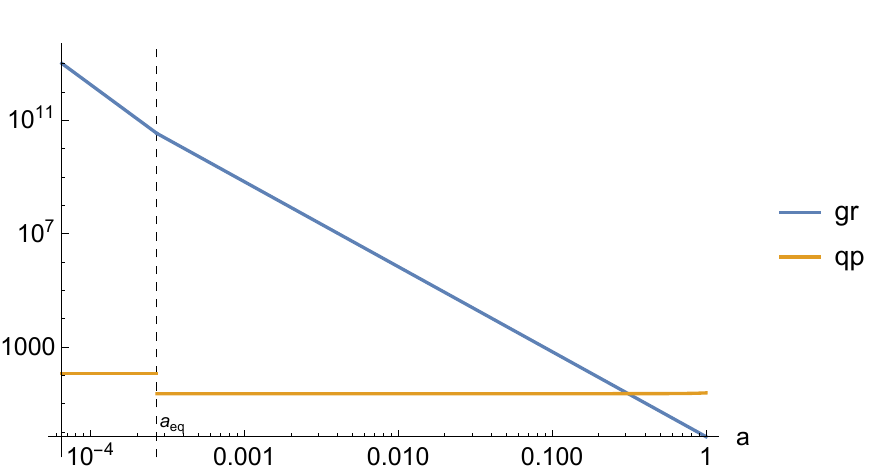} \\
\caption{The comparison of terms for $k/H_0=3\times10^{4.5}$. \\\hspace{200pt}
}
\end{subfigure}\quad
\begin{subfigure}{0.405\textwidth}
	\includegraphics[width=\textwidth]{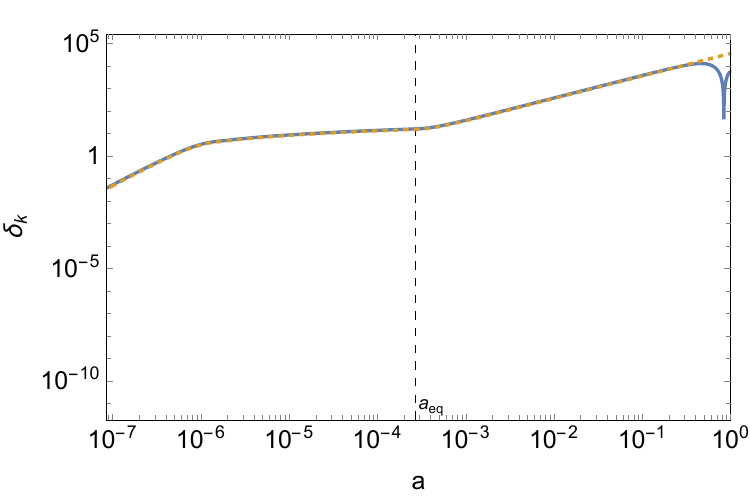}
\caption{$\delta_k$ evolution for our k-ULDM (solid line) and CDM (dashed line), with $k/H_0=3\times10^{4.5}$.}
\end{subfigure}
\\\ \\
\begin{subfigure}{0.46\textwidth}
	\includegraphics[width=\textwidth]{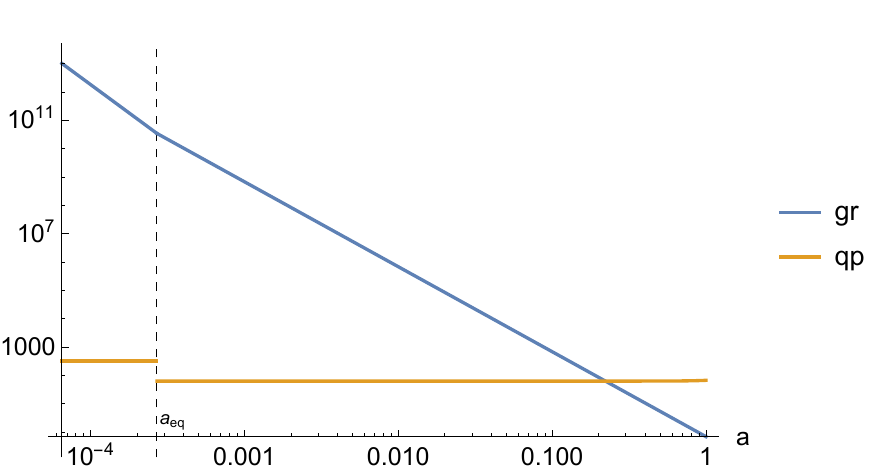} \\
\caption{The comparison of terms for $k/H_0=5\times10^{4.5}$. \\\ 
}
\end{subfigure}
\ 
\begin{subfigure}{0.41\textwidth}
	\includegraphics[width=\textwidth]{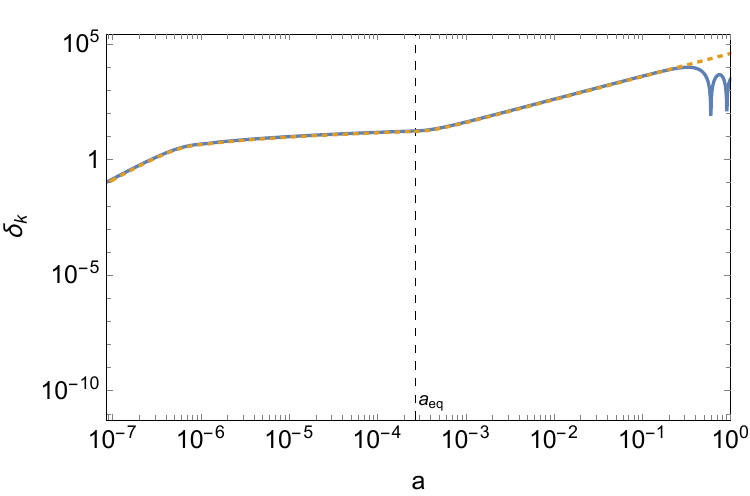}
\caption{$\delta_k$ evolution for our k-ULDM (solid line) and CDM (dashed line), with $k/H_0=5\times10^{4.5}$.}
\end{subfigure}
\caption{\textbf{L}eft panels: The {competition} of gravitational {terms} and sound speed (``quantum pressure'') terms {at scale $k$}. \textbf{R}ight panels: the {corresponding} k-ULDM overdensity evolution, with suppression compared to CDM. {The sound speed term counters the collapse and suppresses the structure formation when it becomes dominant during MD.} {The k-ULDM model here takes} the $f(\phi)$ proposed in \eqref{eqt.fphii}, with $c_{s,i}=c_s(a_i=6.5\times 10^{-5})=1.09\times 10^{-8}$ and $m/H_0=29.0$. }
\label{fig.k3csi}
\end{figure}

\begin{figure}\centering
	\includegraphics[width=0.7\textwidth]{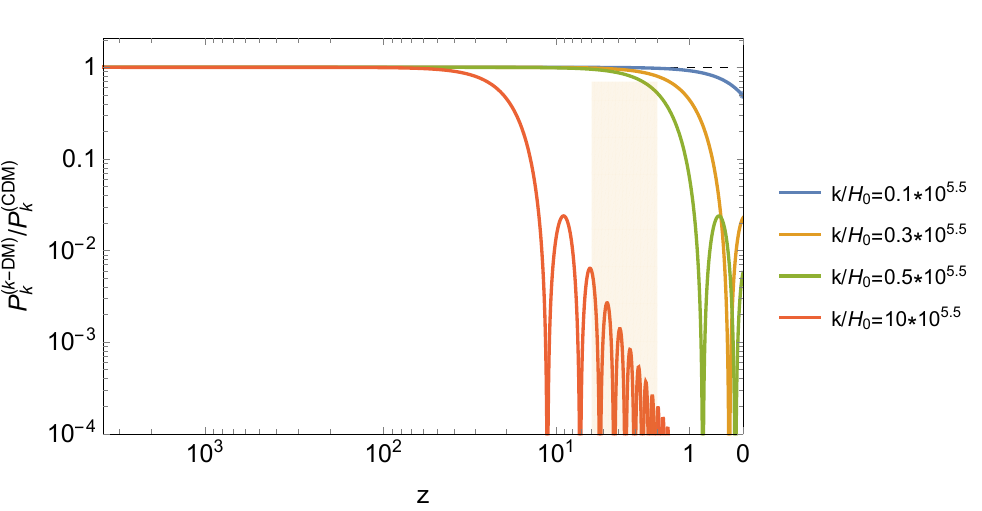}
\caption{{After $t_{\text{eq}}$, }the suppression of overdensity spectrum $P^{(\text{\tiny k-DM})}_k/P^{(\text{\tiny CDM})}_k$ as a function of $z$, for the $f(\phi)$ proposed in \eqref{eqt.fphii} with $c_{s,i}=c_s(a_i=6.5\times 10^{-5})=1.09\times 10^{-8}$ and $m/H_0=29.0$. Again, the isolated blue point at $z=0$ is for {the expected wave nature presented by the k-ULDM, with a half mode }
$k=0.1\times 10^{5.5}H_0\sim 4.5\text{ Mpc}^{-1}$ (corresponding to $10^{-22}$ \text{eV} FDM). The shaded region is the benchmark from the Ly$\alpha$, which should lay below the yellow line for $k=0.3\times 10^{5.5}H_0\sim 20h\text{ Mpc}^{-1}$, as required by \eqref{eqt.lya}. {We can see that such k-ULDM model escape the Ly-$\alpha$ constraints successfully, with a quick suppression at late time. }}
\label{fig.supcsi}
\end{figure}

We depict the overdensity evolution of the k-ULDM from \eqref{eqt.fphii} in figure~\ref{fig.k3csi}. 
The {power spectrum suppression $P^{(\text{\tiny k-DM})}_k/P^{(\text{\tiny CDM})}_k$} and its comparison to {observational benchmarks \eqref{eqt.half}\eqref{eqt.lya}} are shown in figure~\ref{fig.supcsi}. We can see that such k-ULDM model {(corresponding to $10^{-22}$ eV FDM)} escape the Ly-$\alpha$ constraints successfully, with a quick suppression at late time. {Note that the mass of k-ULDM here can be even smaller than in ``phase transition'' case, as $m= 29 H_0\sim 10^{-31}$ eV.
}
\section{Conclusions and outlook}
\indent
We have shown that, by modifying the sound speed with the help of theories with non-canonical kinetics, we can change {the history of structure formation} at small scale. {By delaying small-scale suppression by ULDM wave nature, we can reopen the mass window of ULDM being closed by current or future observations. } We used DBI theory in this paper as an example, and discussed two scenarios for DBI field $\phi$ to be able to serve as DM. One is that when $f(\phi)=0$, the gauge-invariantly defined sound speed $c_s=1$, and DBI scalar goes back to canonical kinetic case, which is similar to FDM. Another is at the {relativistic} limit when $f(\phi)\dot{\phi}^2\rightarrow 1$, so the sound speed is quite small, and the $c_s^{-1}\dot{\phi}^2$ term dominates the energy density. The latter one is in some sense similar to generalized Chaplygin gas without DE constitution. 
\\ \indent
Within the two scenarios, we constructed two exampling models by {assigning} the specific $f(\phi)$. The first is called ``phase transition'' case, making the k-ULDM behaves like CDM at first and like FDM at late time, by switching off $f(\phi)$ after the time $t_c$. This corresponds to transiting from $c_s\ll 1$ scenario {to} $c_s=1$ scenario {of} DBI DM. We have shown a set of parameter values for achieving this, and find that such model can truly reopen the mass window of ULDM, by delaying its small-scale structure suppression. It seems, however, such delayed suppression switched at the time $t_c$ (independent of $k$) is not enough to be compatible {with Ly$\alpha$ benchmark}, due to its sensitivity on $k$ value and also the slow suppression at late time. So we are also interested in another possibility shown in the second example, called ``Chaplygin-like'' case. The DBI DM sticks to the $c_s\ll 1$ scenario, but at late time the $\frac{k^2}{a^2}c_{s,g}^2$ can dominates over gravitational term if we take proper parameters. In this case the delayed small-scale suppression can be integrated more quickly, and the k-dependence is less sensitive, thus better to agree with observations. {Both cases need k-ULDM with lighter masses than {their} corresponding canonical ULDM, pushing it away from BHSR-excluded region.}
\\ \indent
We have just studied few among the wide possibilities of k-ULDM. As can be seen in the appendix \ref{sec.apC}, the transition between two scenarios in DBI can be quite interesting. From this, we can give a model spontaneously transforming from CDM-like to DE-like at late time. {Moreover,} apart from the $f(\phi)$ discussed in paper, there may exist other $f(\phi)$ forms for $c_s$ to vary between $(H/m)^2$ and $1$. Such transitions can be much more complicated, but they are deserved to be explored, even just for {postulating} physical origin of the switched-off $f(\phi)$ in our ``phase transition'' case. Furthermore, non-canonical theories other than DBI remain to be explored, where the $c_s$ evolution have quite rich possibilities to construct models. For example, the theories from modified gravity can make $c_s$ directly dependent on time or scale, not only on field value from $f(\phi)$ as in DBI, making it more flexible.  Besides, small-scale curiosities from other observations \cite{Perivolaropoulos:2021jda,Lovell:2022bhx} remain to be analyzed and possibly be explained using the similar idea of modified $c_s$. Then we may expect some deeper understanding on DM nature implied by observations, from the physical starting points among these theories. 
\\ \indent
There are many works to be done even in the cases presented in this paper. For simplicity to relate theoretical terms to observables, we only analyze the structure growth in linear order, while the overdensity $\delta$ {can} far exceed non-linear order at small scale. Precise results should resort to semi-analytical non-linear methods {\cite{Li:2018kyk, Taruya:2022zmt}} or {even} N-body simulations {\cite{Mina:2020eik, May:2021wwp,May:2022gus}.} 
{
The half-mode fitting of linear power spectrum may not exactly lead to the proper DM halo profiles. However, we can say that our results indicate the expected wave nature. To relate the matter power spectrum to DM halos, we may refer to \cite{Kulkarni:2020pnb,Kawai:2021znq} as a next step.}
Besides, we have not treated e.o.m of fluid in DBI too carefully, as the overdensity $\delta$ in our k-ULDM differs from CDM only at late time during MD epoch. To improve this, a revised axionCAMB is needed, and more possible new effects from the new model are deserved to be discussed. Trying to {figure out} the new physics behind our examples of k-ULDM is also a direction. {{\blue In addition}, due to the concerns on ``fine-tuning'', the initial conditions for ``Chaplygin-like'' case, and the stabilities deserve to be studied in details. Also, for our k-DM to be a realistic DM model, the considerations from {particle} physics {should} be {included}. We will leave these to further works. }
\acknowledgments
{We thank Elisa G. M. Ferreira for supervising this work.} 
We thank Yingying Li, Yi Wang, Xin Ren, Wentao Luo, Masahide Yamaguchi {and Enci Wang} for valuable discussions. 
This work is supported in part by National Key R\&D Program of China (2021YFC2203100, 2024YFC2207500), by NSFC (12261131497), by CAS young interdisciplinary innovation team (JCTD-2022-20), by 111 Project for ``Observational and Theoretical Research on Dark Matter and Dark Energy" (B23042), by Fundamental Research Funds for Central Universities, by CSC Innovation Talent Funds, by USTC Fellowship for International Cooperation, by USTC Research Funds of the Double First-Class Initiative. 
The Kavli IPMU is supported by World Premier International Research Center
Initiative (WPI), MEXT, Japan.
We acknowledge the use of computing facilities of Kavli IPMU, as well as the clusters {\it LINDA} \& {\it JUDY} of the particle cosmology group at USTC.
\appendix
\section{Perturbation theory}\label{sec.apA}
%
Let us go to the perturbation evolution. Recall the FRW metric with first-order perturbation 
\begin{align}
	ds^2=-(1+2\Phi)dt^2+2a(t)B_{,i}dx^idt+a(t)^2[(1-2\Psi)\delta_{ij}+2E_{,i|j}]dx^idx^j~.
\end{align}
For minimally-coupled gravity, we have Einstein equations as 
\begin{align}
	\frac{k^2}{a^2}\Psi_B&=-4\pi G(\delta\rho-3H\delta q)~,\\
	\mathcal{Y}&=-4\pi G \delta q~,\\
	\dot{\mathcal{Y}}+3H\mathcal{Y}+\dot{H}\Phi &=4\pi G(\delta p-\frac{2}{3}k^2\delta\Sigma)~,\\
	\Psi_B-\Phi_B &=8\pi G a^2\delta\Sigma~,
\end{align}
and continuity equation and Euler equation from the conservation $\nabla_\mu T^{\mu\nu}=0$
\begin{align}
	\dot{\delta\rho}+3H(\delta\rho+\delta p)&=\frac{k^2}{a^2}\delta q+(\bar{\rho}+\bar{p})[3\dot{\Psi}+k^2(\dot{E}-B/a)]~,
\label{eqt.cons0}\\
	\dot{\delta q}+3H\delta q&=-\delta p+\frac{2}{3}k^2\delta\Sigma-(\bar{\rho}+\bar{p})\Phi~,
\label{eqt.consi}
\end{align}
where $\Psi_B=\Psi+H\mathcal{X}$ and $\Phi_B=\Phi-\dot{\mathcal{X}}$ are gauge-invariant. We have defined 
\begin{align}
	\mathcal{X}\equiv a^2(\dot{E}-B/a)~,\quad \mathcal{Y}\equiv\dot{\Psi}+H\Phi~,
\end{align}
and matter perturbations from $T_{\mu}^\nu$
\begin{align}
	\begin{aligned}
	&T_0^0=-(\bar\rho+\delta\rho)~, &T_0^i=-(\bar\rho+\bar{p})(v^{,i}-B^{,i})/a~, \\
	&T_i^0=(\bar\rho+\bar{p})a v_{,i}=\delta q_{,i}~, \quad &T_i^j=\delta_i^j(\bar p+\delta p)+\delta\Sigma_{,i}^{,j}~.\hspace{20pt}
	\end{aligned}
\end{align}
The general e.o.m for fluids can be obtained from the conservation equations \eqref{eqt.cons0}\eqref{eqt.consi}, 
\begin{align}\left\{
	\begin{aligned}
	&\dot{\delta}+3H(c_{s,g}^2-w)\delta=-(1+w)(\theta+\dot{h}/2)\\
	&\dot{\theta}+\left[\frac{\dot{w}}{1+w}+(2-3w)H\right]\theta=\frac{k^2}{a^2}\left(\frac{c_{s,g}^2}{1+w}\delta+\Phi\right)
	\end{aligned}~,
\right.
\label{eqt.fluideomA}
\end{align}
where $\delta\equiv \frac{\delta\rho}{\rho}$, $\theta\equiv -\frac{k^2}{a^2}\frac{\delta q}{p+\rho}$, $w\equiv p/\rho$, and gauge-dependent sound speed defined from $c_{s,g}^2\equiv \frac{\delta p}{\delta \rho}$. In Newtonian gauge, at the subhorizon scale $k\gg aH$, we can write \eqref{eqt.fluideomA} {as} 
\begin{align}
	\ddot{\delta}+2H\dot{\delta}+\left(\frac{k^2}{a^2}c_{s,g}^2-4\pi G \bar{\rho}\right)\delta=0~,
\label{eqt.Newtoneom}
\end{align}
where we have used the Poisson equation. 
\\ \indent
{In DBI theory, we can write} $T_{\mu}^{\nu}$ from the action \eqref{eqt.DBIaction}
\begin{align}
	T_{\mu}^{\nu}\equiv-\frac{2g^{\nu\sigma}}{\sqrt{-g}}\frac{\delta(\sqrt{-g}\mathcal{L}_m)}{\delta g^{\mu\sigma}}
=\frac{1}{\sqrt{1-2fX}}\pd_\mu\phi\pd^\nu\phi+\delta_\mu^\nu\left(\frac{1-\sqrt{1-2fX}}{f}-V\right)~,
\end{align}
where $X\equiv-\frac{1}{2}g^{\mu\nu}\pd_\mu\phi\pd_\nu\phi=\frac{1}{2}[(1-2\Phi)\dot{\phi}^2+2\dot{\phi}\delta\dot{\phi}]$ by expanding to the first order. 
We have 
\begin{align}
	\rho+\delta\rho&=\frac{1}{f}\qty(\frac{1}{c_s}-1)+V~,\\
	p+\delta p&=\frac{1}{f}\qty(1-c_s)-V~,
\end{align}
where $f=f(\phi+\delta\phi)$, $V=V(\phi+\delta\phi)$, and $c_s=\sqrt{1-2fX}$ with $X=\frac{1}{2}[(1-2\Phi)\dot{\phi}^2+2\dot{\phi}\delta\dot{\phi}]$. So we can write 
\begin{align}
	\delta\rho&=\frac{1}{f}\left(\frac{1}{c_s}-1\right)\left(-\frac{f'}{f}\delta\phi-\frac{\delta c_s^2}{2c_s^2}\frac{1}{1-c_s}\right)+V'\delta\phi~,\\
	\delta p&=\frac{1}{f}(1-c_s)\left(-\frac{f'}{f}\delta\phi-\frac{\delta c_s^2}{2c_s^2}\frac{c_s}{1-c_s}\right)-V'\delta\phi~,\\
	\delta q&=-\frac{1}{c_s}\dot{\phi}\delta\phi~.
\end{align}
When $c_s\simeq 1$, we have $\frac{1-c_s}{f}\rightarrow X$, and the case turns into scalar field with canonical kinetic term. Take $V=m^2\phi^2/2$, there is 
\begin{align}
	c_{s,g,\text{eff}}^2=\frac{\frac{k^2}{4a^2m^2}}{\frac{k^2}{4a^2m^2}+1}~,
\end{align}
if $t>t_{\text{osc}}$ and $k\gg aH$. When $c_s\rightarrow 0$, we have $|\frac{\delta c_s^2}{2c_s^2}\frac{c_s}{1-c_s}|\gg |\frac{f'}{f}\delta\phi|$, and $|\frac{1}{f}(1-c_s)\left(\frac{\delta c_s^2}{2c_s^2}\frac{c_s}{1-c_s}\right)|\gg |V'\delta \phi|$ even when $H\ll m$, so 
\begin{align}
	c_{s,g}^2\equiv \frac{\delta p}{\delta \rho}=c_s^2~.
\end{align}
With these we can check the structure evolution for the DM constituted by proposed $\phi$ in DBI theory with $f(\phi)$ model.
\section{Structure growth in some toy models of modified $c_s$}\label{sec.apB}
Here we discuss some toy models to see how the modified $c_s$ changes the evolution of $\delta$. Still, we have 
\begin{align}
	\ddot{\delta}_d+2H\dot{\delta}_d+\left(\frac{k^2}{a^2}c_{s,g}^2-4\pi G\rho_d\right)\delta_d=0~,
\end{align}
for $\delta_d\equiv \delta\rho_d/\rho_d$ during MD epoch. For ULDM (or our k-ULDM) constituting all DM, we have $4\pi G\rho_d=\frac{2}{3}t^{-2}$, and $H=\frac{2}{3t}$ during MD, so 
\begin{align}
	\ddot{\delta}_d+\frac{4}{3t}\dot{\delta}_d+\left(\frac{k^2}{a^2}c_{s,g}^2-\frac{2}{3t^2}\right)\delta_d=0~.
\label{eqt.dcseom}
\end{align}
Neglecting the late time DE-domination epoch, we can also write $a=(t/t_0)^{2/3}$, where $t_0$ means the time today. 
\\ \indent
This e.o.m can be solved when the dispersion term is of power-law. When $t^{-2}\gg \frac{k^2}{a^2}c_{s,g}^2$, we have 
$	\delta_d\propto {t^{2/3}}\ \text{or} \ t^{-1}$, the same as in CDM. 
When $\frac{k^2}{a^2}c_{s,g}^2\gg t^{-2}$, for $\frac{k^2}{a^2}c_{s,g}^2=At_0^{-2}\kappa^l (t/t_0)^n$, the solution is of Bessel functions and $|\delta_d|\cdot t^{2/3}\sim f(\kappa^l t^{n+2})$ $(n\neq -2)$, which can be approximated by\footnote{$\kappa$ here is dimensionless momentum defined by $k/t_0^{-1}$.} 
\begin{align}
	\delta_d\sim {t^{-n/4-2/3}}\kappa^{-(l/4)(1+2/n)^{-1}} \text{when } 2\kappa^l t^{n+2}/(n+2)\rightarrow \infty~. 
\end{align}
Or we can say that, for $n<-2$, $\delta_d\propto {t^{-n/4-2/3}}$ at early time, while $\delta_d\propto {t^{2/3}}$ at late time when the gravitational term dominates. This is exactly the case for FDM (with $n=-8/3$). For $n>-2$, $\delta_d\propto {t^{2/3}}$ at early time while $\delta_d\propto {t^{-n/4-2/3}}$ at late time when the sound speed term exceeds the gravitational term.\footnote{The ``late time'' here just describes the period after the term dominance changes. Depending on the specific parameter choices, it can be the late time of MD epoch, or only happens in the future.}
\section{Other forms of modified $c_s$ from DBI}\label{sec.apC}
\begin{figure}[t]\centering
	\includegraphics[width=0.6\textwidth]{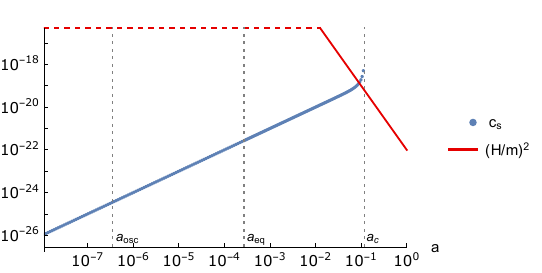}
\caption{An example of $c_s$ growing with time. When $c_s$ grows close to $(H/m)^2$, it jumps to $1$ without hand-set sudden change in $f(\phi)$, then $\phi$ turns back to canonical scalar field. }
\label{fig.csi1}
\end{figure}

Apart from the $c_s\propto t^{-1}$ case as in section~\ref{sec.eg1}, 
we can use 
\begin{align}
	f(\phi)=\left(1-c_{s,i}^2(\phi/\phi_i)^{\frac{4}{\alpha}(\alpha-1+\frac{1}{1+w})}\right)\left(\alpha^{-1} t_i\phi_i^{-1}(\phi/\phi_i)^{\frac{1}{\alpha}-1}\right)^2\quad \text{for } t>t_i \text{\ during RD or MD}~,
\end{align}
to set $\frac{1}{c_s}\dot{\phi}^2\propto a^{-3}$ for $\phi\propto t^\alpha$, and so $c_s\propto t^{2(\alpha-1+\frac{1}{1+w})}$. We need $c_s\ll (H/m)^2$ to make sure $\frac{1}{c_s}\dot{\phi}^2\gg \frac{1}{2}m^2\phi^2$, so that $\rho$ {is dominated by} $\frac{1}{c_s}\dot{\phi}^2$ and such $f(\phi)$ can works well. In this way, the DBI matter can serve as DM (with $p/\rho\simeq c_s\ll 1$) through its kinetic terms only.
\\ \indent
While $(H/m)^2\propto t^{-2}$ for RD or MD, there is some subtlety for the conditions $c_s\ll (H/m)^2$ and $c_s\ll 1$ during the evolution. Let us check \eqref{eqt.conserve}, we have\footnote{When $\alpha+\frac{1}{1+w}=0$, this is the case discussed in section~\ref{sec.eg1}. We have $c_s^{-1}\propto t^{2}\log(At)$, and it holds for $c_s, (H/m)^2\ll 1$.}
\begin{align}
	c_s^{-1}=c_{s,i}^{-1}(t/t_i)^{-2(\alpha-1+\frac{1}{1+w})}-\frac{1}{2\alpha\left(\alpha+\frac{1}{1+w}\right)}
m^2t^2~
\quad (\alpha\neq 0,~\alpha+\frac{1}{1+w}\neq 0)~,
\end{align}
for $c_s< (H/m)^2\ll 1$, and 
\begin{align}
	c_s^{-1}=c_{s,i}^{-1}(t/t_i)^{-2\alpha}+\frac{\alpha^2(\alpha-1)}{\alpha-1+\frac{1}{1+w}}~,
\end{align}
for $c_s< 1\ll (H/m)^2$. By fixing $m$ and $\phi(t)$ while guaranteeing $\phi$ able to serve as DM with a very small $c_s$ for some time, we can see many different behaviors. Besides, the formula indicates that the $c_s$ can jump to $1$ when it grows close to $(H/m)^2$ and then the $\phi$ turns back to canonical scalar field.\footnote{We can get a rough impression of this, noticing that the mass term gets important when $c_s$ grows close to $(H/m)^2$.}
\\ \indent
Here we show an example of $c_s$ growing with time and $m<H_{\text{eq}}$ in figure~\ref{fig.csi1}. 
Without a hand-set sudden change in $f(\phi)$, the $c_s$ will jump to $1$ when it grows close to $(H/m)^2$. Such case can serve as a natural ``phase transition'' from CDM to FDM, with an increasing $c_s$. For an even smaller $m$, we can also have $\phi$ transforming from CDM-like (with small $c_s$) to DE (with $c_s=1$). 
\newpage
\appendix
\bibliography{references}{}
\bibliographystyle{utphys}
\end{document}